\documentclass{aa}

\usepackage{graphicx}
\usepackage{subcaption}
\usepackage{float}
\usepackage{txfonts}
\usepackage[colorlinks=true, allcolors=blue]{hyperref}

\usepackage{xcolor}
\usepackage{booktabs}
\usepackage{tabularx}
\usepackage{placeins}

\begin{document}

   \title{ALMA Chemical Evolution (ACE) survey: The gas fundamental metallicity relation at cosmic noon}

   \authorrunning{I. Langan et al.}          
   \author{I. Langan
          \inst{1},
          I. Shivaei
          \inst{1},
          G. Popping
          \inst{2},
          L. A. Boogaard
          \inst{3},
          N. N. Geesink
          \inst{2},
          M. Kaasinen
          \inst{4},
          A. Pope,
          \inst{5},
          R. Popescu
          \inst{5},
          M. Solimano
          \inst{1},
          L. Arriscado,
          \inst{1},
          D. Narayanan
          \inst{6,7},
          M. Parente
          \inst{6},
          N. Reddy
          \inst{8},
          R. L. Sanders
          \inst{9}}

   \institute{Centro de Astrobiología (CAB), CSIC-INTA, Carretera de Ajalvir km 4, Torrejón de Ardoz, E-28850, Madrid, Spain\\
              \email{imyoko@cab.inta-csic.es}
         \and
             European Southern Observaotry, Karl-Schwartzschild-Str. 2 D-85748, Garching, Germany
         \and
             Leiden Observatory, Leiden University, PO Box 9513, NL-2300 RA Leiden, The Netherlands  
         \and
             Research School of Astronomy and Astrophysics, Australian National University, Canberra, ACT 2611, Australia
         \and
             Department of Astronomy, University of Massachusetts, Amherst, MA 01003, USA
         \and
             Department of Astronomy, University of Florida, 211 Bryant Space Sciences Center, Gainesville, FL 32611 USA
         \and
             Cosmic Dawn Center at the Niels Bohr Institute, University of Copenhagen and DTU-Space, Technical University of Denmark
         \and
             Department of Physics and Astronomy, University of California, Riverside, 900 University Avenue, Riverside, CA 92521, USA   
         \and
            Department of Physics and Astronomy, University of Kentucky, 505 Rose Street, Lexington, KY 40506, USA
 }

   \date{Received September 15, 1996; accepted March 16, 1997}
 
  \abstract   
  {Chemical enrichment shapes how galaxies form and evolve. The gas-phase metallicity is directly linked to the stellar mass, star formation rate, and cold gas of the interstellar medium. Thus, the cold gas fundamental metallicity relation (GFMR) is a powerful tool for probing galaxy evolution, bridging large-scale gas flows modulating the cold gas reservoir and small-scale metal enrichment tracing the cumulative impact of star formation. Constraining all these properties for the same representative sample of galaxies remains challenging yet essential. Using CO(3--2) band 3 observations from the Atacama Large Millimeter/submillimeter Array Chemical Evolution (ACE) survey, we investigated the GFMR in a sample of 26 main-sequence ($\log(\rm M_{*,\rm med})=9.96$), subsolar-metallicity ($12+\log(\rm O/H)_{\rm med}=8.44$) star-forming galaxies (SFGs) at $z\sim2$. With 17/26 CO detections, including some of the lowest-metallicity CO detections at cosmic noon, we find that the stellar mass remains the primary driver of the chemical evolution in our sample ($\sigma_{\rm MZR}\sim0.10$). Whereas the molecular gas likely plays a secondary role ($\sigma_{\rm GFMR}\sim0.11$) similar to that of the star formation rate ($\sigma_{\rm FMR}\sim0.13$). This likely reflects our sensitivity to only the CO-bright component of the molecular reservoir. Our results remain consistent with gas-regulator models and suggest the existence of efficient molecular outflows, with an average mass loading factor of $\eta_{\rm out}\sim4$, regulating star formation and chemical enrichment.}


   \maketitle

\section{Introduction}

The cycling of baryons in and out of galaxies — through gas accretion, star formation, feedback-driven outflows, and gas re-accretion — regulates galaxy evolution across cosmic time (e.g. \citealt{SomervilleDave2015}; \citealt{PerouxHowk2020}; \citealt{Walter2020}).
This baryon cycle thereby sets the gas content, star formation rate (SFR), and chemical enrichment of galaxies and is commonly described by gas-regulator models (e.g. \citealt{Bouche2010}; \citealt{Dave2012}; \citealt{Lilly2013}). Gas-regulator models provide a framework for describing the baryon cycle under which galaxy properties are governed by a quasi-equilibrium between gas inflows, star formation, and gas outflows. In this framework, galaxies self-regulate their growth through the interplay of accretion-driven gas supplies, star formation efficiencies, and feedback-driven losses.

Cosmological simulations and semi-analytic models have demonstrated that gas-regulator processes naturally give rise to tight scaling relations between stellar mass (M$_*$), gas content, the SFR, and gas-phase metallicity\footnote{Throughout the paper, we use metallicity for gas-phase oxygen abundance for readability.} (e.g. \citealt{SomervilleDave2015}; \citealt{Tacchella2016}; \citealt{Torrey2019}; \citealt{MattheeSchaye2019}; \citealt{Dave2020}, \citealt{vanLoon2021}). Observationally, these relations act as tools for probing galaxy evolution and the baryon cycle. The mass–metallicity relation (MZR) shows that more massive galaxies have higher metallicities, reflecting the balance between star formation, subsequent metal production, and metal loss through outflows (e.g. \citealt{Tremonti2004}; \citealt{Maiolino2008}; \citealt{AndrewsMartini2013}; \citealt{Curti2020}; \citealt{Sanders2021}). Extending the MZR, the fundamental metallicity relation (FMR) incorporates the SFR as a third axis -- typically tightening the scatter about the MZR from $\sigma_{\rm MZR}\sim0.1$ to $\sigma_{\rm FMR}\sim0.05$, suggesting that the MZR is not simply redshift-evolving but rather dependent on SFR at fixed M$_*$ (e.g. \citealt{Mannucci2010}; \citealt{LaraLopez2010}; \citealt{Cresci2012}; \citealt{Sanders2015}). Observational results have shown that, at a given M$_*$, the higher the SFR, the lower the metallicity for star-forming galaxies (SFGs). This matches the predictions from gas-regulator models, according to which the metal-poor accreting gas replenishes the gas reservoir of galaxies for subsequent star formation while diluting their metal content (e.g. \citealt{MaiolinoMannucci2019} and references therein; \citealt{Sanders2021}; \citealt{Langan2023}; \citealt{Curti2023}; \citealt{Curti2024}).

Recently, a physically motivated extension of MZRs has emerged: the gas FMR (GFMR). This relation replaces the SFR with molecular gas mass (M$_{\rm mol}$), or gas fraction, establishing a direct link between the metal content of the interstellar medium (ISM) and the gas reservoir that fuels star formation. Observations at low redshifts (e.g. \citealt{Zahid2014}; \citealt{Bothwell2016}; \citealt{Cicone2017}) show that M$_*$, metallicity, and M$_{\rm mol}$ define a tight 3D relation -- tighter than the FMR ($\sigma_{\rm GFMR}\sim0.05$) of those samples, in which gas-rich galaxies are typically more metal-poor at fixed M$_*$. The FMR is thus often interpreted as a byproduct of the GFMR, via the Schmidt-Kennicutt relation (\citealt{Schmidt1959}; \citealt{Kennicutt1998}). In this context, the GFMR provides a more fundamental view of the baryon cycle, as it directly connects the input (gas accretion), processing (star formation), and outputs (metal enrichment and gas outflows) of galaxy evolution.

While the GFMR has been found in the local Universe, observational constraints at high redshifts remain sparse, especially for typical SFGs at cosmic noon (redshift $z\sim2$). This epoch marks the peak of star formation and gas accretion in the universe (e.g. \citealt{MadauDickinson2014}), making it a critical window for testing models of the baryon cycle. However, most high-redshift studies have focused either on the MZR or on molecular gas scaling relations separately, without combining all three GFMR quantities (metallicity, M$_*$, and M$_{\rm mol}$) in the same sample of galaxies. Additionally, existing CO surveys at $z>1$ (e.g. \citealt{Walter2016}, \citealt{Tacconi2018}) have primarily targeted massive, metal-rich systems, leaving the typical star-forming population -- particularly subsolar galaxies -- largely unexplored.

In this paper, we present new constraints on the GFMR at cosmic noon using a sample of intermediate-metallicity galaxies with CO(3--2) observations from the Atacama Large Millimeter/submillimeter Array (ALMA) Chemical Evolution (ACE) survey (\citealt{Shivaei+}). Our targets span a range in M$_*$ and metallicities extending current molecular gas studies at high redshifts to lower-mass, less chemically evolved systems (see Sect.~\ref{subsec:sample}). By combining rest-frame optical spectroscopy and multi-wavelength photometry with our ALMA observations, we simultaneously constrained M$_*$, metallicity, and M$_{\rm mol}$ for each galaxy in our sample. This enabled us to directly test the GFMR in a previously unexplored parameter space at $z\sim2$ and assess whether the relation observed at low redshifts persists in the early universe and for less evolved galaxies.

The paper is structured as follows. In Sect.~\ref{section:two}, we briefly describe our observations and ancillary data used in this paper, as well as the methodology used to derive the CO(3--2) integrated fluxes. In Sect.~\ref{section:three}, we show how we derived M$_{\rm mol}$ and the results on the GFMR at $z\sim2$. In Sect.~\ref{section:four}, we discuss our findings in context of the literature and models. We summarise our conclusions in Sect.~\ref{section:five}. We assumed the \citet{Chabrier2003} initial mass function (IMF) and adopted $12+\log \rm O/H = 8.69$ as the solar oxygen abundance (\citealt{Asplund2009}). The cosmology assumed throughout this work follows the flat $\Lambda$ cold dark matter standard cosmological parameters: $H_0 = 67.66\, \text{km}\, \text{s}^{-1}\, \text{Mpc}^{-1}$, $\Omega_m = 0.311$, and $\Omega_\Lambda = 0.689$ (\citealt{Planck2020}).

\section{Sample, observations and data analysis}
\label{section:two}
In this section we detail the methodology used to measure the CO(3--2) integrated fluxes of the galaxies in our ACE sample. We refer the reader to the ACE survey paper (\citealt{Shivaei+}) and subsequent studies (e.g. \citealt{Popescu+}; \citealt{Geesink+}) for further data reduction not detailed in this paper (e.g. stacking techniques).

\subsection{Sample selection and characteristics}
\label{subsec:sample}
We used observations of the 25, $z=2.1-2.5$ galaxies observed under the ALMA large programme ACE (\#2024.1.00534.L). The ACE sample is a subset of the MOSFIRE Deep Evolution Field (MOSDEF) survey (\citealt{Kriek2015}), selected based on detections of multiple strong rest-frame optical emission lines, ALMA band 6 measurements (\citealt{Shivaei2022}) and free of active galactic nucleus (AGN) signs based on X-rays, IR, and [NII]/H$\alpha$ criteria. Of the 27 galaxies in \citet{Shivaei2022}, 25 were selected for ALMA band 3 and band 7 observations, based on robust metallicity estimates and clear 1.2 mm dust-continuum detections. The metallicities were derived from multiple robust rest-frame optical emission lines (e.g. [OII]$\lambda\lambda$ 3726, 3729, and [SII]$\lambda\lambda$ 6716, 6731 from \citealt{Kriek2015}). We defer to the ACE survey paper (\citealt{Shivaei+}) for more details on the sample selection. Here, we used the ALMA band 3 CO(3--2) data to derive the cold molecular gas masses. To this aim, we used the 22 ACE galaxies targeted in ALMA band 3 (the remaining 3 galaxies being only targeted in ALMA band 7) and an additional 4 galaxies with ALMA band 3 CO(3--2) detections from \citet{Sanders2023} (hereafter S23), also drawn from MOSDEF. This resulted in a total primary sample of 26 galaxies.

The ACE sample lies in the Cosmic Evolution Survey (COSMOS) field (\citealt{Scoville2007}). We gathered the available multi-wavelength photometry available (e.g. from the 3D-HST survey \citealt{Skelton2014}) and included new \textit{\emph{James Webb}} Space Telescope (JWST) NIRCam and MIRI photometry (COSMOS2025: \citealt{Casey2023}; \citealt{Shuntov2025}; PRIMER: \citealt{Dunlop2021}). With this archival photometric data and the new ACE ALMA band 7 and band 3 continuum measurements, we derived M$_*$ from spectral energy distribution (SED) fitting with \texttt{Prospector} (\citealt{Leja2017}; \citealt{Johnson2021}). We updated the existing metallicity measurements based on multiple rest-frame optical emission lines, using the strong line calibrations presented in \citet{Sanders2025}. The SFRs are based on the dust-corrected H$\alpha$ luminosities, as in \citet{Reddy2015} and \citet{Shivaei2016}.  We defer to \citet{Shivaei+} for the exact details on the methodology used to derive M$_*$, metallicity, and the SFRs for our ACE sample, as well as for the added literature objects with available data. Our primary sample of 26 galaxies has M$_*$ values ranging from $\log(\rm M_*/M_\odot)=9.13-10.62$ ($\log(\rm M_*/M_\odot)_{\rm med}=9.96$), SFRs from log(SFR$/M_\odot$yr$^{-1}$) $=1.3-2.2$ ($\log(\rm SFR/M_\odot\rm yr^{-1})_{\rm med} = 1.84$) and metallicities ranging from 12$+\log(\rm O/H)=8.32-8.63$ (12$+\log(\rm O/H)_{\rm med}=8.44$). We report these properties in Table~\ref{table:pties}.

Additionally, we complemented our ACE sample with literature data at similar redshifts, including information on molecular gas (from CO(3--2) or lower-J transition data), metallicity (from rest-frame optical emission lines), SFRs, and M$_*$, and no evidence of hosting an AGN. We included seven galaxies observed under the ALMA Spectroscopic Survey in the Hubble Ultra Deep Field (ASPECS; \citealt{Walter2016}; \citealt{Aravena2019}; \citealt{GonzalezLopez2019}; \citealt{Boogaard2019}; \citealt{Boogaard2020}) and five galaxies observed under the Plateau de Bure High-z Blue Sequence Survey (PHIBSS; \citealt{Tacconi2013}; \citealt{Tacconi2018}). To ensure a self-consistent analysis, we computed the metallicity, M$_*$, and M$_{\rm mol}$ of the added literature data using the same methodology (see \citealt{Shivaei+} and Sect.~\ref{subsec:cotoMmol}).  We report these properties in Table~\ref{table:ptiesliterature}. For the remainder of this paper, we use the extended sample made of our primary ACE sample and the complementary literature data.

\subsection{ALMA CO(3--2) observations}
\label{subsec:observations}

S23 present a subset of the MOSFIRE survey with follow-up ALMA band 3 observations (\#2018.1.01128.S), resulting in six detections, two tentative detections, and six non-detections. All six galaxies in the S23 sample that are not secure detections (i.e. the tentative and non-detections) were also observed with ACE (\citealt{Shivaei+}). For galaxies observed in both ACE and S23, we combined the two measurement sets to the same frequency grid with CASA \texttt{tclean} using the coarser spectral resolution. We retained the combined data cubes only when the average root mean square (rms) of the spectrum (see Sect.~\ref{subsubsec:spectra}) improved by more than 10\%. Two galaxies (IDs 3666 and 24763) met this criterion. For these, we used the combined S23+ACE data. For the remaining four galaxies, we only used the ACE observations.

\subsubsection{Data reduction: CO(3--2) data cubes}
\label{subsubsec:ACEdatareduction}
We used the Common Astronomy Software Applications (CASA 6.5.21) data processing software \citep{CASA2022} to perform the data reduction (\texttt{tclean}) in a homogenised manner on all galaxies of our sample. We imaged the CO(3--2) (hereafter CO\footnote{If not specified CO refers to CO(3--2) throughout this paper.}) data cubes with natural weighting ($\texttt{robust}=2$) and twice the native resolution,\footnote{We tested three times the native resolution and find that the best balance between optimising the $S/N$ and preserving the line shape is achieved with two-channel averaging.} resulting in channel width $\Delta v=7.8125$ MHz (or $\Delta v\sim23\, \text{km\,s}^{-1}$). 

The band 3 observations were taken at different times during cycle 11, with different antenna configurations; therefore, the beam resolution varies across our sample. As a result, we find that 9 out of 26 galaxies are marginally resolved and require tapering to ensure that all CO(3–2) flux is captured within a single beam element (see Sect.~\ref{subsubsec:spectra} for details on flux measurements). To assess the varying resolution, we imaged each galaxy at its native resolution and with additional 1$\arcsec$ and 2$\arcsec$ tapering, resulting in resolutions of $\sim1.5\arcsec$ (native) to $\sim3\arcsec$ (2$\arcsec$ tapering) for the galaxies with the highest resolution. For each source, we extracted CO spectra at the peak pixel of the moment-zero map from the native-resolution, 1$\arcsec$, and 2$\arcsec$ tapered data cubes. To determine the optimal cube for flux measurement, we computed a score for each cube, defined as the product of its integrated line flux and signal-to-noise ratio ($S/N$), each normalised by the maximum flux and $S/N$ among the three cubes at different spatial resolutions for that source. This enabled us to keep a balance between sensitivity and capturing all CO emission. The cube with the highest score was selected as the preferred resolution for reporting the final integrated flux.

For sources where a tapered cube (1$\arcsec$ or 2$\arcsec$) was selected, we further verified the choice using a JvM-corrected curve of growth analysis (\citealt{JvM1995}) with the package \texttt{interferopy}\footnote{https://github.com/interferopy/interferopy} (\citealt{interferopy}). Specifically, we verified that the curve peaked at or near the corresponding beam size, consistent with the assumption that spectra extracted from the peak pixel approximate the flux within one synthesised beam. In addition to this quantitative selection procedure, we visually inspected the moment-zero maps and complementary JWST imaging to assess potential contamination from nearby sources. In the case of galaxy ID 4497 (see Table~\ref{table:COfluxes}), although the scoring method indicates that the 2$\arcsec$ tapered data yielded the highest score, inspection of the moment-zero map and JWST images reveals the presence of a neighbouring source within the larger beam. Consequently, the apparent increase in flux in the tapered data is likely due to contamination from this nearby source rather than to additional flux from the target galaxy itself. In this specific case, we chose the 1$\arcsec$ tapered cube as the final cube. In non-detections cases (see Sect.~\ref{subsubsec:spectra}), we kept the native data cube as the final cube.

Once all CO data cubes were imaged, we created optimised moment-zero maps and spectra. 1) We used the observed H$\alpha$ full width at half maximum ($\pm$FWHM) from MOSFIRE observations around the system redshift to create a first moment-zero map. 2) We identified the peak of the emission of that moment-zero map and used this position to extract the CO spectrum from the cube. 3) We fitted this spectrum (see Sect.~\ref{subsubsec:spectra}) and used the best-fit FWHM to create the optimised moment-zero map. 4) We identified the peak of the emission of that optimised moment-zero map and used this position to extract the final optimised CO spectrum from the cube. As a final check, we compared the integrated flux (see Sect.~\ref{subsubsec:spectra}) from the optimised CO spectrum and the peak flux from the optimised moment-zero map and find that those are fully consistent within the errors. In the non-detection cases, the moment-zero map was computed as in step 1), and the spectrum was extracted from a centred beam-like aperture. In $\sim65$\% of our sample (17/26), CO is detected at $>3 \, \sigma$.

\subsubsection{Emission line fitting: CO(3--2) integrated flux}
\label{subsubsec:spectra}
To measure the CO integrated line fluxes, we applied the same Monte Carlo (MC) fitting procedure to all CO spectra. For each spectrum, we first measured the average rms noise (rms$_{av}$), and used it to generate 5000 noisy realisations of the observed spectrum. We produced each realisation by perturbing the spectrum with Gaussian noise drawn from a distribution centred on rms$_{av}$. Then, we fitted each perturbed spectrum with a single Gaussian profile, requiring the amplitude ($A$) and width ($\sigma$) to be positive while leaving the remaining parameters free. From each fit, we computed the integrated flux assuming a Gaussian profile, i.e. $F_{int} = A \times \sigma \times \sqrt{2\pi}$. The final integrated flux is the median of all 5000 realisations (column 3 in Table~\ref{table:COfluxes}). We also took the median of the parameters fit as the final best-fit parameters, from which we find consistent integrated fluxes. We report those best-fit parameters in Table~\ref{table:Fitparams}. In Fig.~\ref{fig:COHAFWHM}, we compare the resulting CO and H$\alpha$ FWHMs, highlighting potential differences in the kinematics of the different gas phases.

For low $S/Ns$, defined as $3\leq S/N<4$ on the integrated flux, we repeated the MC procedure described above but by constraining the Gaussian width using the observed FWHM of the H$\alpha$ emission line (FWHM$_{\rm H\alpha}$). We adopted a prior centred on the FWHM$_{\rm H\alpha}$, with a distribution width that accounts for the combined spectral resolution of the H$\alpha$ and CO observations. This applies to 10 out of the 26 sources (marked as Fit $=0$ in Table~\ref{table:COfluxes}).

The uncertainty on the final integrated flux is taken as the larger of (i) the standard deviation of the 5000 integrated flux measurements and (ii) the noise-based estimate, rms$_{av} \times \sqrt{N_{chan}} \times \Delta v$,  where $N_{chan}$ is the number of channels within $\sigma$ of the median Gaussian and $\Delta v$ is the spectral resolution. This conservative choice prevents artificially small uncertainties in cases where the Gaussian width is constrained and the MC dispersion is therefore intrinsically reduced. We note that typically (i) is the maximum.

\begin{figure*}
    \centering
    \includegraphics[width=0.9\textwidth]{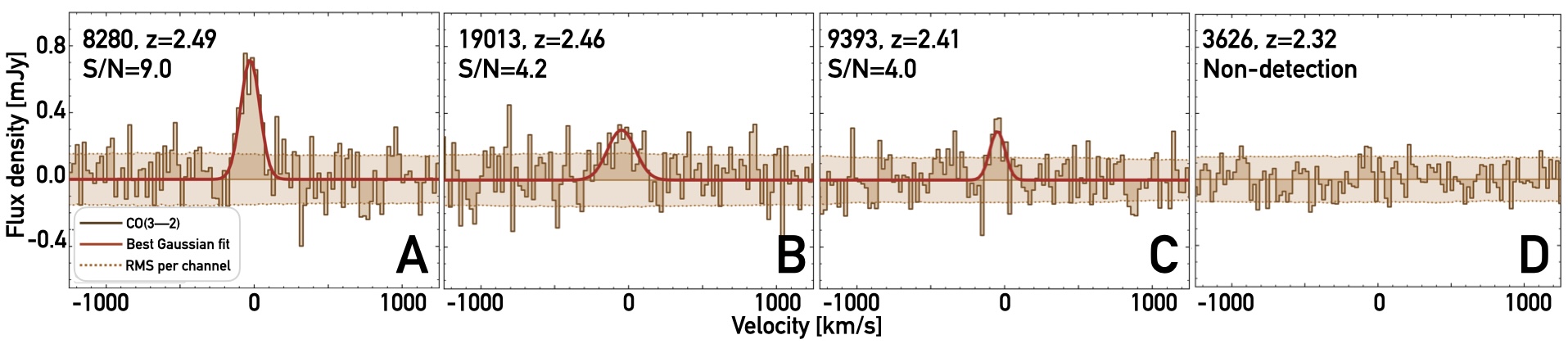}
    \caption{ACE CO(3--2) spectrum examples (light-brown shaded area). We show the rms per channel, with the horizontal beige-shaded area within the dotted lines and the Gaussian fit (red line) used to measure the integrated CO(3--2) flux. A) and B): Our highest and lowest $S/N$ detections, respectively, fitted with the line width as a free parameter. C): Case where we constrained the fit with the FWHM$_{\rm H\alpha}$. D): Non-detection where the spectrum is extracted from a centred beam-like aperture.}
    \label{fig:COspectrum_example}
\end{figure*}

We defined non-detections as integrated fluxes with $S/N<3$. In such cases, we set $3 \sigma$ upper limits as $3\times rms_{av}$ under $\pm$ FWHM$_{\rm H\alpha}$ around the systemic redshift, i.e. $3 \times rms_{av} \times \sqrt{N_{chan}} \times \Delta v$, where $N_{chan}$ is the number of channels under 2 FWHM$_{\rm H\alpha}$. The CO integrated fluxes used in this paper are presented in Table~\ref{table:COfluxes}, and Fig.~\ref{fig:COspectrum_example} shows four examples illustrating the methodology followed in this section. In total, we have 17 detections with an average $S/N=5$ and nine upper limits. In Fig.~\ref{fig:spectra} we show the full sample of CO spectra, minus the four examples in Fig.~\ref{fig:COspectrum_example}.

\begin{table*}[ht!]
\caption{CO-derived measurements for our ACE sample.}
\label{table:COfluxes}
\centering
\begin{tabular}{|cccccccc|}
\hline
\toprule
\textbf{ID} & \textbf{logM($_{\rm mol}$)} & \textbf{S$\Delta v_{\text{CO}(3-2)}$} & \textbf{L$'_{\text{CO}(1-0)}$} & \textbf{Method} & \textbf{Beam} & \textbf{Fit} & \textbf{Data}\\
(1) & (2) & (3) & (4) & (5) & (6) & (7) & (8)\\
\hline
\midrule
\vspace{1pt}
2672 &  $10.91_{-0.013}^{+0.14}$ & $145 \pm 34$ & $5.2 \pm 1.3$ & tap2as & 3.2 & 1 & S23\\

3324 &  $10.40_{-0.11}^{+0.11}$ & $86\pm 17$ & $3.2 \pm 0.6$ & tap2as & 3.0 & 1 & ACE\\

3626$^\dag$  & $<10.56$ & $<33$ & $<1.2$ & native & 2.2 & 1 & ACE\\

3666  & $10.52_{-0.10}^{+0.10}$ & $44\pm 10$ & $1.4 \pm 0.3$ & native & 2.9 & 0 & ACE+S23\\

3773$^\dag$ &  $<10.57$ & $<38$ & $<1.5$ & native &1.6 & 1 & ACE \\

4497 &  $10.42_{-0.11}^{+0.11}$ & $68\pm15$ & $2.8 \pm 0.6$ & tap1as &2.0 & 1 & ACE\\

5094 &  $10.84_{-0.13}^{+0.12}$ & $191\pm41$ & $6.3 \pm 1.4$ & tap1as &1.7 & 0 & S23\\

5814 & $10.67_{-0.13}^{+0.12}$ & $119\pm20$ & $3.8 \pm 0.6$ & native &3.1 & 1 & ACE\\

5901$^\dag$ &  $<10.25$ & $<37$ & $<1.4$ & native &1.4 & 0 & ACE\\

6283$^\dag$ & $<10.57$ & $<37$ & $<1.3$ & native &2.0 & 1 & ACE\\

6750$^\dag$ & $<10.31$ & $<35$ & $<1.1$ & native &1.4 & 1 & ACE\\

8280 & $10.65_{-0.10}^{+0.09}$ & $116\pm 13$ & $4.9 \pm 0.6$ & native &2.3 & 1 & ACE\\

8515$^\dag$ & $<10.64$ & $<41$ & $<1.7$ & native &1.6 & 1 & ACE\\

9393 &$10.41_{-0.12}^{+0.11}$ & $39\pm 10$ & $1.5 \pm 0.4$ & native &1.4 & 0 & ACE\\

9971 &$10.70_{-0.11}^{+0.11}$ & $76\pm 19$ & $3.0 \pm 0.8$ & tap2as &3.0 & 0 & ACE\\

13296 & $10.26_{-0.12}^{+0.11}$ & $88\pm 18$ & $2.9 \pm 0.6$ & tap1as &2.2 & 1 & S23\\

13701 & $10.52_{-0.10}^{+0.09}$ & $116\pm 22$ & $3.8 \pm 0.7$ & tap2as &2.9 & 1 & S23\\

16594 & $10.43_{-0.10}^{+0.10}$ & $63\pm 13$ & $2.3 \pm 0.5$ & tap1as &2.1 & 0 & ACE\\

19013 & $10.35_{-0.12}^{+0.12}$ & $58\pm 14$ & $2.4 \pm 0.6$ & native &1.5 & 0 & ACE\\

19439$^\dag$ &$<10.59$ & $<32$ & $<1.3$ & native &1.6 & 1 & ACE \\

%20661 &19753 &  2.4694& $10.94_{-0.03}^{+0.04}$ & $158\pm 34$ & $6.5 \pm 1.4$ & tap1as &2.0 & 1 & S23\\

19985 &  $10.63_{-0.10}^{+0.10}$ & $60\pm13$ & $2.0 \pm 0.4$ & native &2.2 & 1 & ACE\\

21955 &  $10.46_{-0.11}^{+0.11}$ & $67\pm 16$ & $2.8 \pm 0.7$ & native &1.6 & 0 & ACE\\

22193$^\dag$ &  $<10.25$ & $<34$ & $<1.4$ & native &1.5 & 1 & ACE\\
 
24020$^\dag$ & $<10.29$ & $<33$ & $<1.0$ & native &2.1 & 1 & ACE\\

24763 & $10.55_{-0.12}^{+0.12}$ & $67\pm13$& $2.7 \pm 0.5$ & native &1.8 & 0 & ACE+S23\\

25229 & $10.45_{-0.14}^{+0.14}$ & $41\pm12$ & $1.4 \pm 0.4$ & native &1.9 & 0 & ACE\\[3pt]

\hline
\bottomrule
\end{tabular}
\tablefoot{(1) lists the IDs from the 3D-HST v4 catalogue (\citealt{Skelton2014}). (2) lists the M$_{\rm mol}$ in solar masses, derived using the \citet{Accurso2017} metallicity-dependent $\alpha_{\rm CO}$. (3) lists the CO(3--2) integrated flux in milijansky kilometre per second. (4) lists the CO(1--0) luminosity in [$10^9$ K km$s^{-1}$ pc$^2$]. It is the conversion from CO(3--2) integrated flux, given the excitation conversion factor $r_{31}=0.77$ from \citet{Boogaard2020}. (5) lists the spatial resolution used when imaging the data cube. (6) lists the average beam resolution in [arc seconds] of the final data cube used to measure the CO(3--2) integrated flux. (7) indicates whether the Gaussian fit of the CO(3--2) spectrum is free (`1') or constrained to the FWHM$_{\rm H\alpha}$ (`0'). (8) indicates which observations are used, i.e. ACE only, S23 only, or a combination of both. We report 3$\sigma$ upper limits for sources marked with $^\dag$.}
\end{table*}

\section{Results}
\label{section:three}
\subsection{CO flux to molecular gas mass}
\label{subsec:cotoMmol}
To derive M$_{\rm mol}$ from CO integrated flux measurements we need to assume a CO-to-H$_2$ conversion function (i.e. $\alpha_{\rm CO}=M_{\rm mol}/L_{\text{CO}(1-0)}=r_{31}M_{\rm mol}/L_{\text{CO}(3-2)}$). Several studies have shown evidence of a metallicity-dependent $\alpha_{\rm CO}$ (e.g. \citealt{Bolatto2013b}, \citealt{Madden2020}, S23); therefore, we used our rare combination of metallicity and CO observations at $z\sim2$ to implement a metallicity-dependent $\alpha_{\rm CO}$.

\begin{figure*}[]
    \centering
    \begin{subfigure}[b]{0.48\textwidth}
         \centering
         \includegraphics[width=0.9\textwidth]{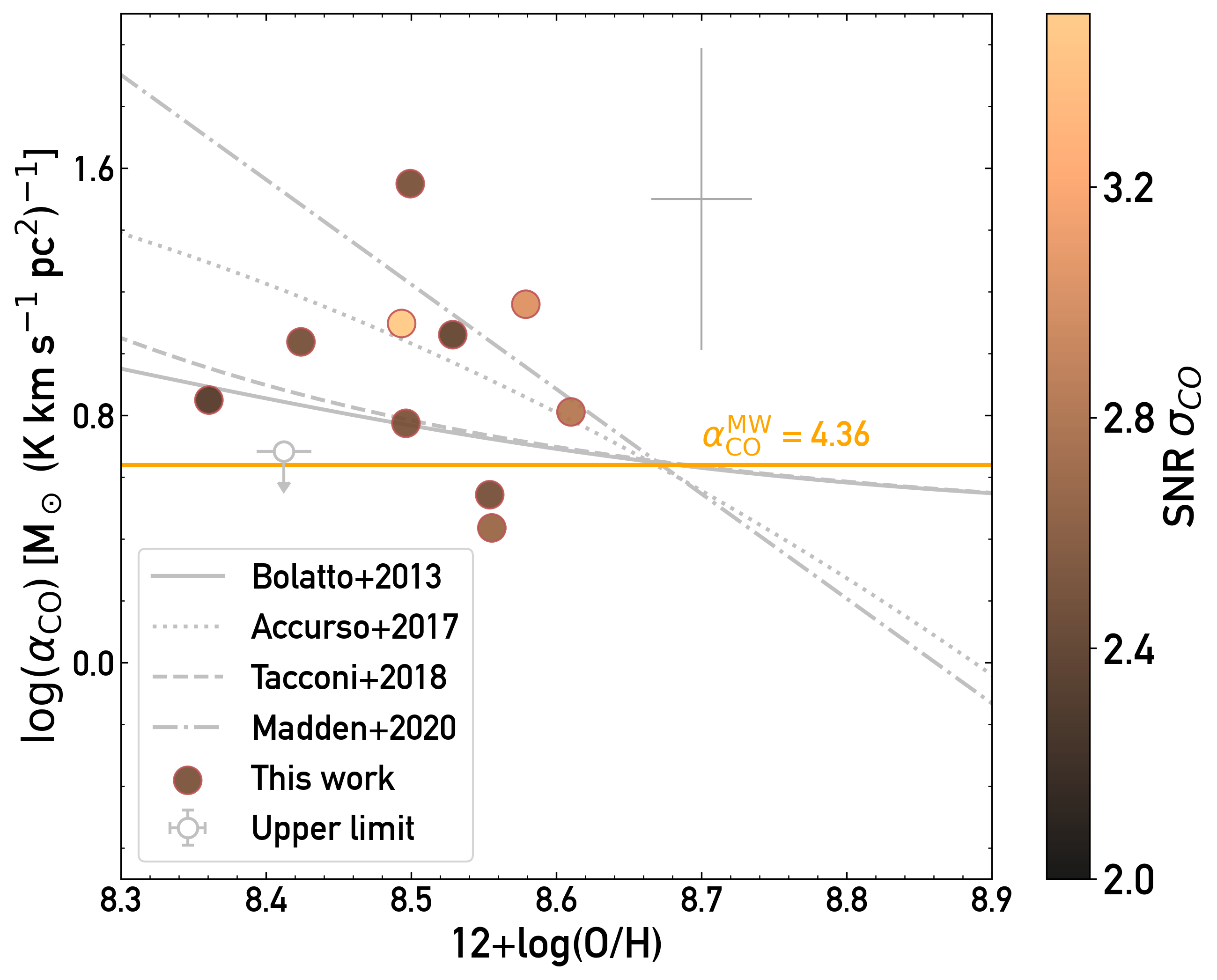}
     \end{subfigure}
     \begin{subfigure}[b]{0.48\textwidth}
         \centering
         \includegraphics[width=0.9\textwidth]{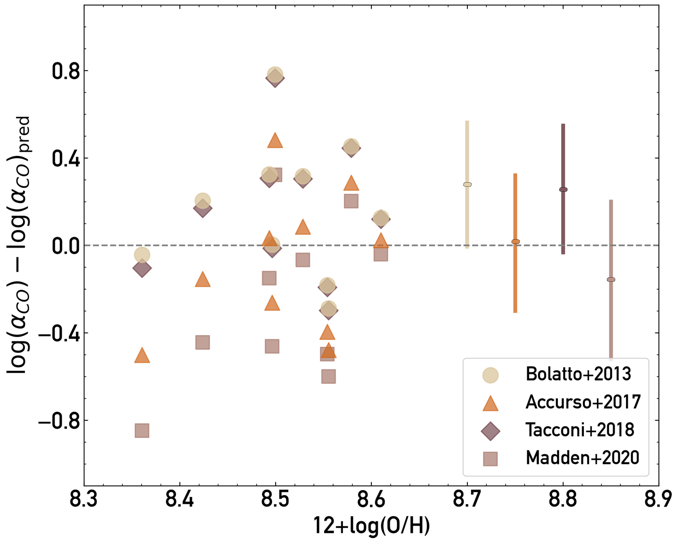}
     \end{subfigure}
\caption{Left: CO-to-H$_2$ conversion function, $\alpha_{\text{CO}}$, as a function of metallicity, $12+\log(\text{O/H})$. Our measured $\alpha_{\text{CO}}$ values, derived using the dynamical method, are shown with circles, colour-coded by the $S/N$ of the CO line width. The median uncertainties on both axes are shown with the grey error bars on the top right. Different values of $\alpha_{\text{CO}}$ from the literature are shown with solid (\citealt{Bolatto2013b}), dotted (\citealt{Accurso2017}), dashed (\citealt{Tacconi2018}), and dash-dotted (\citealt{Madden2020}) grey lines. The horizontal beige line shows the constant Milky-Way value, $\alpha_{\text{CO}}^{\text{MW}}=4.36$. Right: $\alpha_{\text{CO}}$ residuals, i.e. our measured $\alpha_{\text{CO}}$ minus the predicted one, for different $\alpha_{\text{CO}}$ calibrations as a function of metallicity. The residuals for different $\alpha_{\text{CO}}$ calibrations are shown with different markers. The vertical bars in matching colours placed at arbitrary metallicities show the weighted residuals scatter and median. Based on this we chose $\alpha_{\text{CO}}$ from \citet{Accurso2017} for this paper.}
\label{fig:alphaCO}
\end{figure*}

\paragraph{The dynamical mass method.} To make an educated decision on which metallicity-dependent $\alpha_{\rm CO}$ to choose from the literature, we estimated our own metallicity-dependent $\alpha_{\rm CO}$ using the dynamical mass method (see e.g. S23) on our CO detected galaxies. The dynamical mass method assumes that $z\sim2$ SFGs are baryon-dominated within their effective radii, i.e. dark matter is negligible (e.g. \citealt{Genzel2017}, \citealt{Genzel2020}, \citealt{Price2021}), and that the total gas mass is dominated by molecular gas (e.g. \citealt{Tacconi2018}). Under these assumptions, we estimated the molecular gas mass as $M_{\rm mol} \simeq M_{\rm dyn} - M_*$. The dynamical mass was computed as $M_{\rm dyn} = k R_{\text{eff}} \sigma_v^2 G^{-1}$, where $G$ is the gravitational constant, $R_{\text{eff}}$ is the effective radius, $k$ is the virial coefficient, and $\sigma_v$ is the integrated line-of-sight CO velocity, obtained from a Gaussian fit to the unresolved spectrum and corrected for instrumental spectral resolution. We adopted $R_{\text{eff}}$ as provided by COSMOS2025 (\citealt{Shuntov2025}) based on a multi-band model-fitting approach using \texttt{SourceXtractor++} (\citealt{Bertin2020}, \citealt{Kummel2020}, \citealt{Kummel2022}), and we measured $\sigma_v$ from the CO emission line widths (corrected for the resolution of our ALMA observations). We adopted $k=8.6$ as in S23, where $k$ was empirically calibrated by matching dispersion-based dynamical masses to rotation-based estimates for galaxies with resolved kinematics. This resulting value is therefore an average value, recovering the dynamical mass on for a population with random inclinations. As in S23, we did not apply individual inclination correction due to unresolved CO data and the lack of resolved rotation measurements. We therefore note that our $M_{\rm dyn}$ estimates may be affected by individual unknown inclinations. For the purpose of this subsection only, we applied the procedure explained in Sect.~\ref{subsubsec:spectra} to find the best-fit $\sigma_v$ to all 17 CO detections in the same way. That is, unlike in Sect.~\ref{subsubsec:spectra} where we discussed constraining the width of the fit for low $S/N$ cases, here we left it free. We individually cross-checked the best-fit parameters for those low $S/N$ cases where the fit was left free to vary or was constrained to the FWHM$_{\rm H\alpha}$ and find good agreement. We assumed a CO(3--2)/CO(1--0) luminosity ratio of $r_{31}=0.77\pm0.14$, as inferred by \citet{Boogaard2020} for similar SFGs at the same redshift as the ACE sample ($z\sim2-3$). To compare with theoretical predictions, we also computed $r_{31}$, as in \citet{Narayanan2014} for unresolved observations, and find, on average, $r_{31}\sim0.84$, consistent with our adopted value $r_{31}=0.77\pm0.14$. Galaxies with IDs 19985, 8280, and 25229 have M$_{dyn}<\rm M_*$, which would result in an unphysical negative M$_{\rm mol}$; therefore, these galaxies were excluded the analysis presented in this section. We also removed galaxies with a $S/N$ of their best-fit width, $\sigma_v$, lower than 2. We show the result of our metallicity-dependent $\alpha_{\rm CO}$ using the dynamical mass method in Fig.~\ref{fig:alphaCO} (left). The median error bar is shown in the upper right of the plot. Given the large uncertainty on our metallicity-dependent $\alpha_{\text{CO}}$ data points (mainly due to the uncertainty on the CO emission line widths), we cannot provide a meaningful metallicity-dependent $\alpha_{\text{CO}}$ fit with higher accuracy than what already exists in the literature.

\paragraph{Metallicity-dependent $\alpha_{\text{CO}}$ from the literature.} Several metallicity-dependent $\alpha_{\text{CO}}$ have been proposed. We focused on four of these, spanning a wide range in $\alpha_{\text{CO}}$. First, we adopted the values recommended for $z\sim1-2$ main-sequence (MS) SFGs and/or for subsolar SFGs from \citet{Bolatto2013b}, based on earlier theoretical work from \citet{Narayanan2012}:

\begin{equation}
    \alpha_{\rm CO} = 0.67 \times \exp{(0.36 \times 10^{-Z_g})},
\end{equation}

where $Z_g=12+\log \rm O/H - 8.69$. Second, we take $\alpha_{\text{CO}}$ from \citet{Accurso2017}, calibrated on observations of subsolar SFGs with $\log(\rm M_*/M_\odot)=9-10$ and radiative transfer modelling:

\begin{equation}
    \alpha_{\rm CO} = 10^{14.752 - 1.623\times (Z_g+8.69)} + 0.062\times \log(\Delta \rm MS),
\label{eq:alphaCO}
\end{equation}

where $\Delta \rm MS=\rm SFR/SFR_{MS}$ is the offset from MS (\citealt{Shivaei2015}). Third, we adopted the $\alpha_{\text{CO}}$ from \citet{Tacconi2018}, which was computed for the PHIBSS sample of $z=0-4$ galaxies with $\log(\rm M_*/M_\odot)=9.0-11.8$ and $-1.3<\log(\Delta \rm MS) <2.2$:

\begin{equation}
\alpha_{\rm CO} = 4.36 \times r_{J1} \times \sqrt{0.67\exp(0.36 \times 10^{-Z_g}) \times 10^{-1.27 Z_g}},
\end{equation}

where $Z_g=12+\log \rm O/H - 8.69$, and $r_{J1}=r_{31}=0.77$ in our case (\citealt{Boogaard2020}).

Fourth, we adopted the $\alpha_{\text{CO}}$ from \citet{Madden2020}, which probes CO-dark molecular gas resulting from low-metallicity environments down to $1/50$ solar and find 

\begin{equation}
    \alpha_{\rm CO} = 10^{0.58 - 3.39\times(12+\log \rm O/H-8.69)}.
\end{equation}

In the right panel of Fig.~\ref{fig:alphaCO}, we show the residuals of $\log(\alpha_{\text{CO}})$, i.e. the difference between our $\alpha_{\rm CO}$ inferred from the dynamical method and the different literature $\alpha_{\rm CO}$ calibrations, corrected to the same metallicity scale (i.e. \citealt{Sanders2025}) and normalised to the solar value $\alpha_{\rm CO}^{\rm MW}=4.36$. Residuals are represented using different symbols: beige circles for \citet{Bolatto2013b}, orange triangles for \citet{Accurso2017}, dark-brown diamonds for \citet{Tacconi2018}, and rosy-brown squares for \citet{Madden2020}. For each calibration, vertical bars indicate the weighted (i.e. based on the uncertainties in our measured $\alpha_{\text{CO}}$) mean and scatter, in matching colours. We then computed a score for each calibration, defined by the combination of weighted means and scatters, and identified the model yielding the lowest score as the most consistent with our data. This scoring approach was designed to identify the calibration that yields residuals closest to zero (minimising bias, i.e. maximising accuracy) and with the smallest scatter (i.e. optimising precision). The \citet{Accurso2017} and \citet{Tacconi2018} relations yield the best (i.e. lowest) scores. We also computed the reduced $\chi^2$ between each calibration and our data. We find that the \citet{Bolatto2013b} and \citet{Accurso2017} relations yield the best reduced $\chi^2$. Therefore, for the remainder of our analysis, we adopted the metallicity-dependent $\alpha_{\rm CO}$ relation from \citet{Accurso2017}, Eq.~\ref{eq:alphaCO}, which includes a 1.36 correction factor accounting for helium and metals. We used the MS presented in \citet{Shivaei2015}, corrected for the updated M$_*$ computed with ACE. \citet{Accurso2017} specify the validity domain of this prescription is limited to SFGs with $7.9 <12+\log(\rm O/H) <8.8$ and $-0.8 <\log (\Delta\rm MS) <1.3$\footnote{Our sample has a mean $\Delta(\log$MS$_{\rm av})= 0.5$ dex}. In Table~\ref{table:COfluxes}, we show the resulting M$_{\rm mol}$.

\subsection{Gas fundamental metallicity relation (GFMR)}
\label{subsec:GFMR}
We fitted the relation between metallicity, M$_*$, and M$_{\rm mol}$ (hereafter GFMR) using Bayesian statistical modelling and MCMC implemented in \texttt{PyMC} (\citealt{Patil2010}). The metallicity, $y=12+\log(\rm O/H)$, was modelled as
 
\begin{equation}
y = a\log \left( \frac{\rm M_*}{\rm M_{*,0}} \right) + b \log\left(\frac{\rm M_{\mathrm{mol}}}{\rm M_{\mathrm{mol},0}}\right) + c,
\end{equation}

where $\log(\rm M_{*,0})=10.28$ and $\log(\rm M_{\mathrm{mol},0})=10.57$ are the mean values of the stellar and detected molecular masses of our extended sample, respectively. M$_*$ and M$_{\rm mol}$ were each modelled as normal distributions centred on their observed values, with standard deviations given by the average of the asymmetric measurement uncertainties. In the specific case of molecular gas non-detections, i.e. upper limits, we applied a censored likelihood using a cumulative distribution function, corresponding to the probability that the true value lies below the detection threshold. We adopted a weakly informative normal prior on the true M$_{\rm mol}$, with mean equal to the sample average and standard deviation of 1 dex. The observed metallicities were also modelled as normally distributed around the linear relation, with a variance given by the measurement uncertainty combined in quadrature with an intrinsic scatter term. We adopted broad normal priors for the slope parameters $a\sim \mathcal{N}(0.3,1)$ and $b\sim \mathcal{N}(0.08,1)$, as well as the intercept $c\sim \mathcal{N}(6.53,1)$, based on the results of \citet{Bothwell2016}. The posterior is sampled using the No-U-Turn Sampler (NUTS) with four chains, 1000 tuning steps, and 2000 draws per chain.
This framework allowed us to jointly fit the scaling relation while propagating measurement uncertainties and incorporating upper limits.

Based on the posterior which we show in Fig.~\ref{fig:cornerplot}, we find the following GFMR:

\begin{equation}
    y = 0.19^{+0.03}_{-0.03}\log\left(\frac{\rm M_*}{\rm M_{*,0}}\right) -0.03^{+0.04}_{-0.04}\log\left(\frac{\rm M_{\mathrm{mol}}}{\rm M_{\mathrm{mol},0}}\right)\\
    + 8.51^{+0.02}_{-0.01}
\label{eq:GFMR}
\end{equation}

with the intrinsic scatter in $y$ around the best-fit plane $\sigma_{\rm GFMR} = 0.11_{-0.01}^{+0.02}$ dex. The intrinsic scatter of the GFMR is a parameter of the model we fit, which accounts for observational uncertainties.

\begin{figure*}[]
    \centering
    \includegraphics[width=\textwidth]{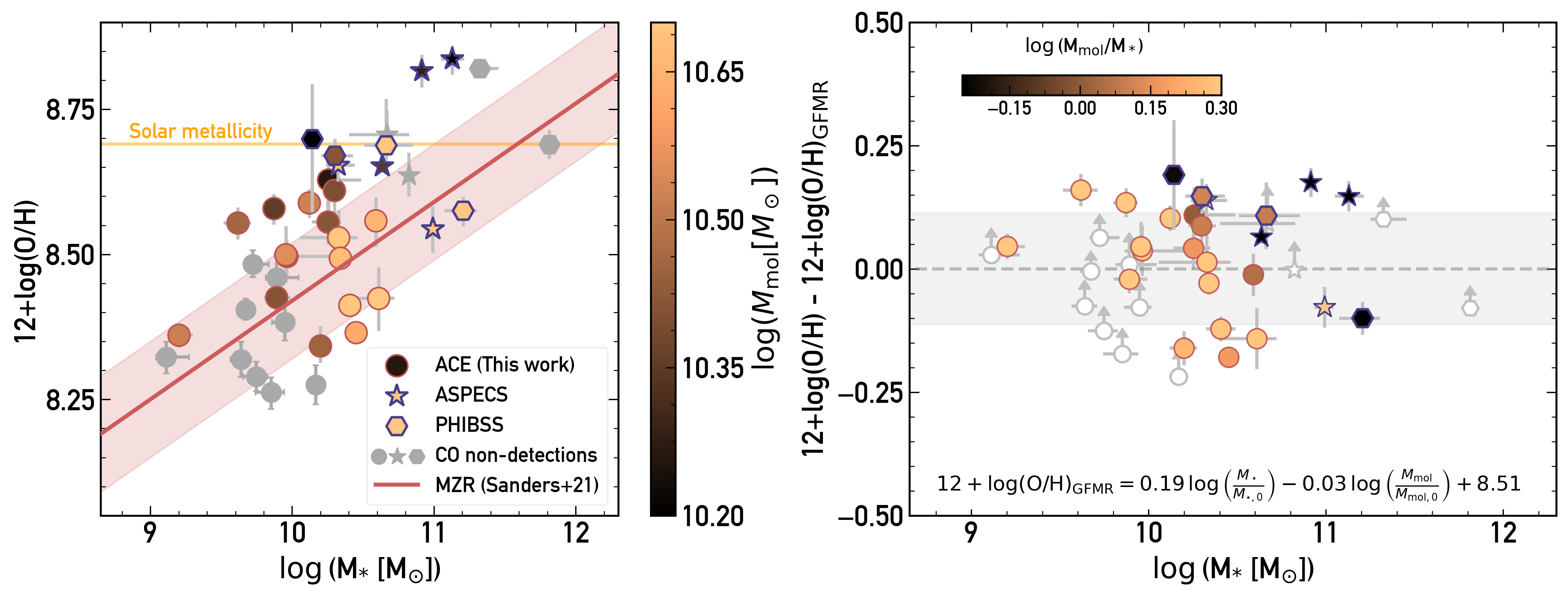}
\caption{Left: GFMR, i.e. metallicity as a function of stellar mass (M$_*$), colour-coded by molecular gas mass (M$_{\rm mol}$). CO non-detections are shown with grey markers. As a reference we show the solar metallicity value with the horizontal orange line and the MZR from \citet{Sanders2021} with the red line. Right: Residuals, i.e. the measured metallicity minus its prediction given our best GFMR fit (Eq.~\ref{eq:GFMR}), as a function of M$_*$, colour-coded by the molecular gas-to-stellar mass ratio. This best GFMR fit is shown at the bottom of the plot, and its intrinsic scatter is measured to be $\sigma_{\rm GFMR}\sim0.11$ (grey-shaded region). CO non-detections correspond to lower limits in the residuals and are shown with grey arrows pointing upwards. In both figures the stars represent ASPECS galaxies, and the hexagons represent PHIBSS galaxies.}
\label{fig:GFMR_GFMRres}
\end{figure*}

We show the results in Fig.~\ref{fig:GFMR_GFMRres}, i.e. the metallicity ($12 + \log(\text{O/H})$) as a function of the M$_*$ of our sample, colour-coded by the M$_{\rm mol}$ where CO non-detections are shown by grey circle markers. The left panel shows the MZR colour-coded with M$_{\rm mol}$. The right panel shows the residuals of the GFMR -- i.e. the difference between our measured metallicity and the predicted one according to our best-fit GFMR -- colour-coded by the molecular gas-to-stellar mass ratio (M$_{\rm mol}/\rm M_*$). The M$_{\rm mol}$ non-detections are shown with grey arrows pointing upwards, as they are upper limits in M$_{\rm mol}$. In both panels, the ASPECS and PHIBSS galaxies are shown with stars and hexagons, respectively. 

\section{Discussion}
\label{section:four}
\subsection{The relation between molecular gas mass and metallicity}
\label{subsec:mmol}
\begin{figure*}[]
    \centering
    \includegraphics[width=0.85\textwidth]{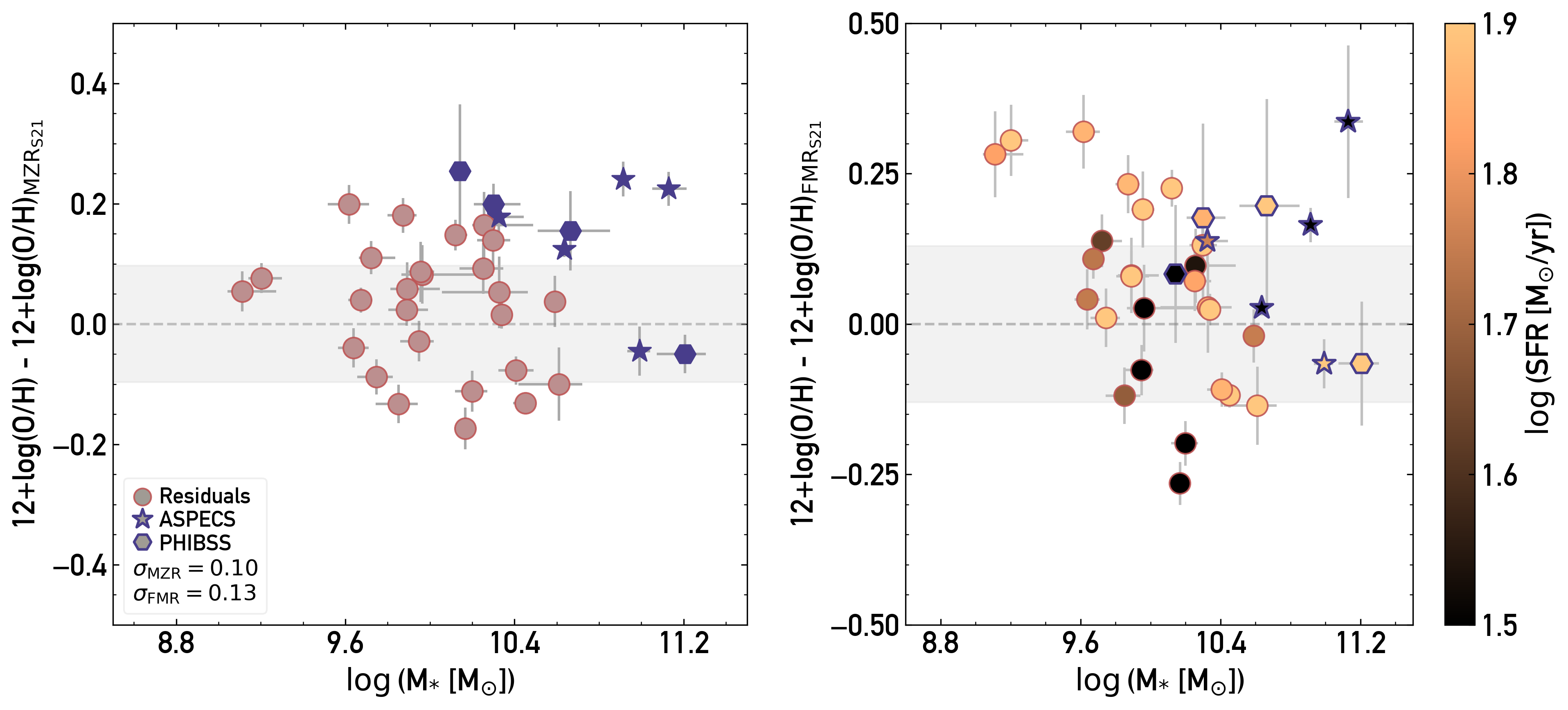}
\caption{MZR (left) and FMR (right) residuals, i.e. our measured metallicity minus the predicted one given the MZR$_{\rm S21}$ (FMR$_{\rm S21}$) for the parent sample of ACE. Both MZR$_{\rm S21}$ and FMR$_{\rm S21}$ are corrected according to the same metallicity scale (\citealt{Sanders2025}) used in this paper. The intrinsic scatter of the residuals is represented by the horizontal shaded region, $\sigma_{\rm MZR}\sim0.10$ for MZR$_{\rm S21}$ and $\sigma_{\rm FMR}\sim0.13$ for FMR$_{\rm S21}$. On both plots, the stars represent ASPECS galaxies, and the hexagons represent PHIBSS galaxies.}
\label{fig:res}
\end{figure*}

To investigate how the molecular gas content impacts the metallicity, we first examined the residuals of the MZR, shown in the left panel of Fig.~\ref{fig:res}. These residuals are defined as the difference between the observed metallicity and those predicted by the MZR derived for the parent sample from which our ACE sample was drawn (MZR$_{\rm S21}$; \citealt{Sanders2021}). We converted MZR$_{\rm S21}$ to the same metallicity calibration adopted in this work (\citealt{Sanders2025}) to account for systematic offsets between calibrations. The residuals are well distributed around zero, with an intrinsic scatter of $\sigma_{\rm MZR}\sim0.10$ dex (after accounting for observational uncertainties), indicating that the MZR$_{\rm S21}$ provides an adequate description of our data. This is expected, given that ACE is a subsample of the parent dataset used to compute MZR$_{\rm S21}$ and that the MZR is a well-established relation across cosmic time (e.g. \citealt{MaiolinoMannucci2019}).

We then extended the MZR by including the molecular gas mass, M$_{\rm mol}$, and fitted a linear relation between metallicity, stellar mass, and molecular gas mass (see Sect.~\ref{subsec:GFMR}). This yields a weak dependence on M$_{\rm mol}$ ($b=-0.02\pm0.04$), much weaker than the dependence on M$_*$ ($a=0.19\pm0.03$), and an intrinsic scatter of $\sigma_{\rm GFMR}\sim0.11$ dex. While the residuals are centred around zero, the scatter does not decrease compared to the MZR, suggesting that M$_{\rm mol}$ does not provide independent predictive power beyond stellar mass. However, we note that M$_{\rm mol}$ is derived from CO observations using a metallicity-dependent $\alpha_{\text{CO}}$. As a result, the weak dependence on M$_{\rm mol}$ and the lack of improvement in intrinsic scatter can be caused by the introduced coupling between M$_{\rm mol}$ and metallicity. To mitigate this effect, instead of M$_{\rm mol}$, we repeated the analysis using the observed CO luminosity, L$'_{\text{CO}(1-0)}$ (Table~\ref{table:COfluxes}). In this case, we find a weak dependence on L$'_{\text{CO}(1-0)}$ ($b=0.03\pm0.01$), and an intrinsic scatter of $\sigma\sim0.12$ dex. The scatter remains similar to that of the MZR, indicating that the inclusion of CO luminosity does not provide independent and additional predictive power. The detected correlation is therefore best interpreted as a secondary effect. Physically, these observations likely reflects the known dependence of CO luminosity on metallicity, whereby higher-metallicity environments exhibit more efficient dust shielding and thus brighter CO emission (e.g. \citealt{Stacey1991}, \citealt{Wolfire2010}, \citealt{Accurso2017}, \citealt{Madden2020}). In low-metallicity environments, CO emission becomes faint or undetectable, where molecular hydrogen gas can survive while CO is photodissociated. This CO-dark molecular gas may therefore constitute a substantial fraction of the total H$_2$ reservoir. In this sense, L$'_{\text{CO}(1-0)}$ traces the CO-bright molecular gas rather than the total molecular gas reservoir. Therefore, the lack of a strong correlation between CO-bright molecular gas and metallicity may be biased rather than a lack of direct causal link between the total molecular gas and chemical enrichment, as discussed in \citet{Bothwell2013}. Total M$_{\rm mol}$ estimates are needed to further explore the link between metallicity, M$_*$, and gas mass.

For comparison, we also examined the FMR, shown in the right panel of Fig.~\ref{fig:res}, using the parametrisation from \citet{Sanders2021} (FMR$_{\rm S21}$). The residuals are similarly centred around zero, with an intrinsic scatter of $\sigma_{\rm FMR}\sim0.13$ dex, slightly larger than that of the MZR. This modest increase is likely due to the limited dynamic range in SFR probed by our sample. This suggests that stellar mass remains the primary parameter governing the chemical state of galaxies in our sample. We do not find significant evidence of additional dependence on SFR or CO-bright molecular gas, although the limited dynamic range and sample size prevent us from placing strong constraints on such potential secondary trends.

Although we fitted a three-parameter relation involving M$_*$, M$_{\rm mol}$, and metallicity, given the lack of improvement in predictive power and the potential biases associated with $\alpha_{\text{CO}}$, we caution against interpreting this as evidence of the fundamental role of molecular gas in setting metallicity. Instead, our results indicate that any such dependence is subdominant and sensitive to how the molecular gas content is inferred.

Finally, we explored the role of molecular gas fraction by examining trends with $\mu_{\rm gas}=\rm M_{\rm mol}/M_*$ (Fig.~\ref{fig:GFMR_GFMRres}, right panel). We observe a mild anti-correlation between metallicity and gas fraction at fixed stellar mass, consistent with a scenario in which galaxies with larger gas reservoirs relative to their stellar mass have converted a smaller fraction of their baryons into stars and are therefore less chemically enriched. Recent results, for example from \citet{Tacchella2023} and \citet{LangeroodiHjorth2023}, who analysed galaxy morphology with respect to their metallicity and locus on the FMR at high redshift, found that more compact objects are preferentially less metal-rich. This points towards rapid gas accretion on timescales shorter than those associated with ISM enrichment. However, observational constraints on chemical evolution and gas accretion from quasar absorption studies (e.g. \citealt{Langan2023}) find that gas dilution and star formation-driven enrichment are closely coupled (i.e. in the FMR). These results argue against a simple temporal delay between rapid inflow-driven dilution and subsequent chemical enrichment -- although we note that O enrichment occurs very rapidly, so any temporal delay is very short and probably not easily captured by the observations. Instead, they favour a picture in which gas accretion, star formation, and metal enrichment operate in a quasi-equilibrium manner, i.e. as in the gas-regulator framework. In this context, galaxies with higher $\mu_{\rm gas}$ have lower metallicities because they have processed a smaller fraction of their gas into stars, not because enrichment lags significantly behind gas accretion. Thus, it is not the absolute molecular gas mass but rather the balance between gas and stellar mass that governs the observed metallicity.

\subsection{The GFMR in the framework of gas-regulator models}
Gas-regulator models describe how gas inflows and outflows jointly regulate the chemical and overall evolution of SFGs by controlling the gas reservoir available for star formation and by redistributing metals between the ISM and the circumgalactic medium (e.g. \citealt{SomervilleDave2015}; \citealt{Veilleux2020}; \citealt{Ginolfi2020a}; \citealt{Ginolfi2020b}; \citealt{Freundlich2021}; \citealt{Wendt2021}; \citealt{Langan2023}). In this framework, galaxies evolve towards a quasi-equilibrium state in which gas accretion, star formation, and outflows are closely coupled. Within the gas-regulator formalism (e.g. \citealt{Lilly2013}), the evolution of the metallicity in the ISM can be written as

\begin{equation}
    Z_{\text{ISM}}=Z_0 + \frac{y}{1 + \eta_{out}(1-R)^{-1} + f_{\rm gas}},
\label{eq:nout}
\end{equation} where $Z_0$ is the metallicity of the accreting gas (often assumed to be negligible compared to $Z_{\rm ISM}$), $f_{\rm gas}$ is the gas fraction, $R$ is the return fraction (how much M$_*$ is returned to the ISM), $y$ is the stellar yield (how much newly formed metals are released into the ISM), and $\eta_{out}$ is the outflow mass loading factor (efficiency with which gas outflows can remove mass and metals from galaxies), defined as $\eta_{out}=\dot{\rm M}_{out}/\mathrm{SFR}$.

\begin{figure}[]
    \centering
    \includegraphics[width=0.85\columnwidth]{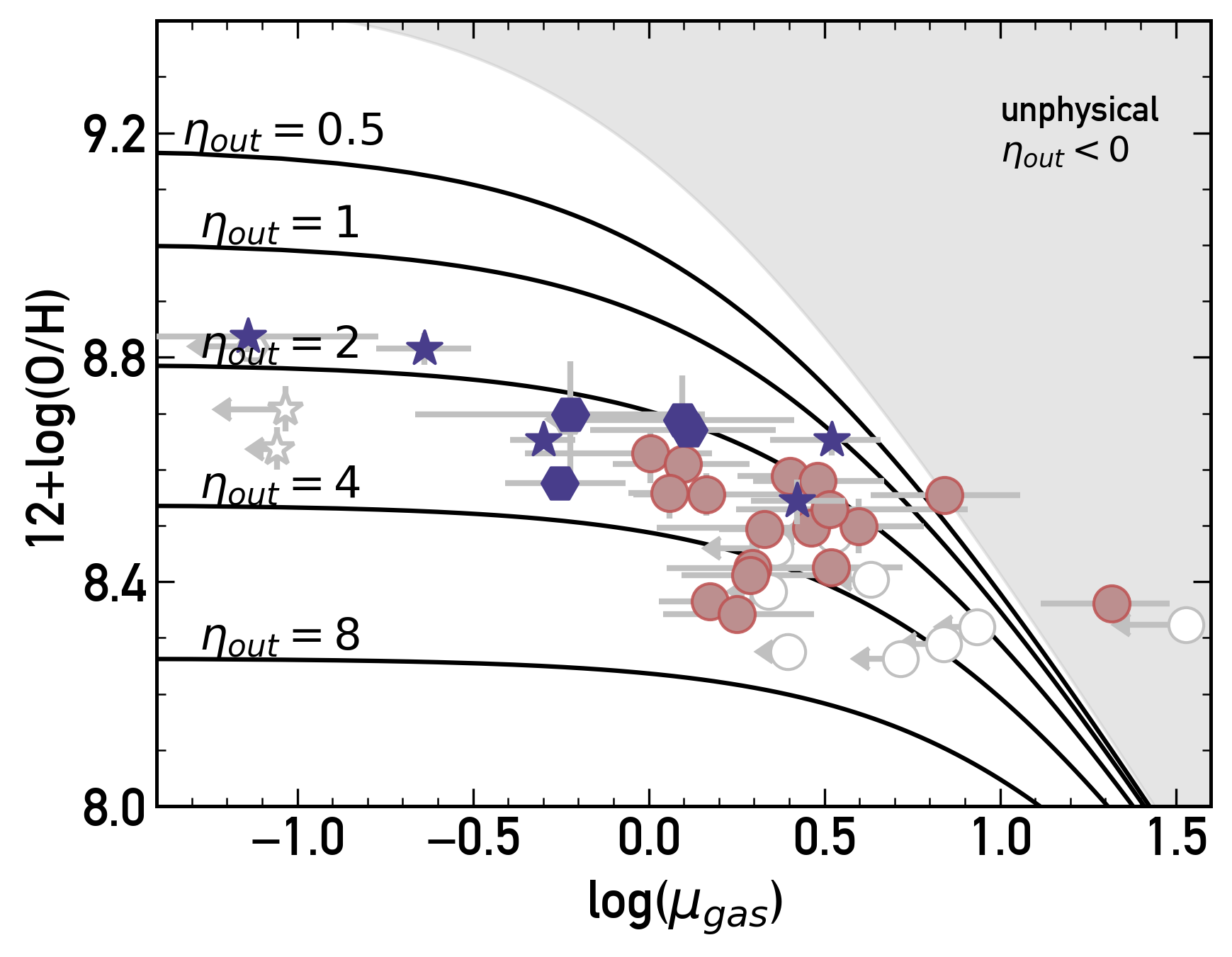}
\caption{Metallicity vs the molecular gas-to-stellar ratio defined as $\mu_{\rm gas}=\rm M_{mol}/\rm M_*$. The individual ACE galaxy detections are shown with red circles, while the non-detections are shown with grey arrows. The purple markers show literature data from ASPECS (star) and PHIBSS (hexagons). The black lines show the theoretical predictions from the gas-regulator model of \citet{Lilly2013}, with outflow mass-loading factors ranging from $\eta=0.5$ to $8$. The unphysical parameter space ($\eta<0$) is shown with the grey-shaded area.} 
\label{fig:nout}
\end{figure}

In Fig.~\ref{fig:nout}, we show theoretical predictions (solid black lines) of the outflow mass loading factor $\eta_{out}$ as a function of $\mu_{\rm gas}$ (Eq.~\ref{eq:nout}), i.e. we assume the total gas mass is dominated by the molecular gas and use $\mu_{\rm gas}$ as a proxy for the gas fraction. We take $R=0.45$ and $y=0.0333$ as appropriate for core-collapse supernova enrichment with a \citet{Chabrier2003} IMF (\citealt{Vincenzo2016}). We over-plot the metallicities of our ACE sample as a function of $\mu_{\rm gas}$ (red circles, white circles with an arrow for upper limits in M$_{\rm mol}$, and brown squares for binned values). Our lowest-metallicity galaxy, ID 3666, lies in the unphysical $\eta_{\rm out}<0$ region (grey-shaded area). This may reflect a slight overestimate of $M_{\rm mol}$ (given its low $S/N$ ), an incorrect assumed $\alpha_{\rm CO}$ for this specific system, an underestimate of $M_\star$, or point to a physically more complex system (e.g. merger, clumpy), as suggested by its rest-frame optical morphology (\citealt{Shivaei+}). We find an average $\eta_{out} \sim 4$ for our sample, with values decreasing from $\eta_{out}\sim6$ to $\eta_{out}\sim1$ as $\mu_{\rm gas}$ increases. This trend reflects the anti-correlation between metallicity and gas fraction discussed in Sect.~\ref{subsec:mmol} and is qualitatively consistent with expectations from gas-regulator models, where systems with lower gas fractions are more chemically enriched. Indeed, simulations (e.g. \citealt{Langan2020}) predict that, at a fixed stellar mass, metallicity increases as the gas fraction decreases, in agreement with the overall trend observed in Fig.~\ref{fig:GFMR_GFMRres}, right panel. Furthermore, at fixed $\mu_{\rm gas}$, we observe that more metal-rich systems correspond to lower inferred values of $\eta_{out}$. This suggests that more massive galaxies (via the MZR) experience less efficient outflows. This is consistent with a scenario in which deeper gravitational potential wells inhibit the removal of enriched material, leading to higher metal retention efficiencies.

The inferred average value of $\eta_{out} \sim 4$ is consistent with S23 and \citet{Barfety2025} (in their $z>1.7$ and averaged $\log(M_*/M_\odot)=10.8$ sample) and suggests that molecular outflows are efficient at removing mass from galaxies across stellar masses and metallicities. Within the gas-regulator framework, such values imply that molecular outflows strongly regulate both star formation and chemical enrichment, unlike ionised outflows for which $\eta_{out}$ is typically found to be $\leq1$ (e.g. \citealt{Freeman2019}, \citealt{ForsterSchreiber2019}, \citealt{Weldon2024}), even in galaxies hosting an AGN (e.g. \citealt{Davies2020}, \citealt{Bertola2025}, \citealt{Venturi2025}). This difference suggests that the bulk of the mass outflow may reside in cooler gas phases. This interpretation is supported by recent hydrodynamical simulations (e.g. \citealt{Chen2026}), which show that the cool ($T<10^4$ K) phase dominates the mass loading of outflows in SFGs at cosmic noon, carrying the majority of the mass out of galaxies.

\subsection{Comparison with results in the literature}
The GFMR was first proposed by \citet{Bothwell2016}, presenting an analysis combining various samples in the local Universe, such as the APEX Low-redshift Legacy Survey for MOlecular Gas (ALLSMOG; \citealt{Bothwell2014}), the CO Legacy Database for the GASS survey (COLD GASS; \citealt{Saintonge2011}) and \textit{Herschel} Reference Survey (HRS; \citealt{Boselli2010}), with information on metallicity, M$_{\rm mol}$, and M$_*$. One of the main results of this study is the following linear relationship defining the metallicity as a function of M$_{\rm mol}$ and M$_*$ ($\rm GFMR_{B16}$):

\begin{equation}
    12+\log(\rm O/H) = 0.31\log(\rm M_*) - 0.08\log(\rm M_{mol}) +6.53.
\end{equation}

\begin{figure}
    \centering
    \includegraphics[width=0.9\columnwidth]{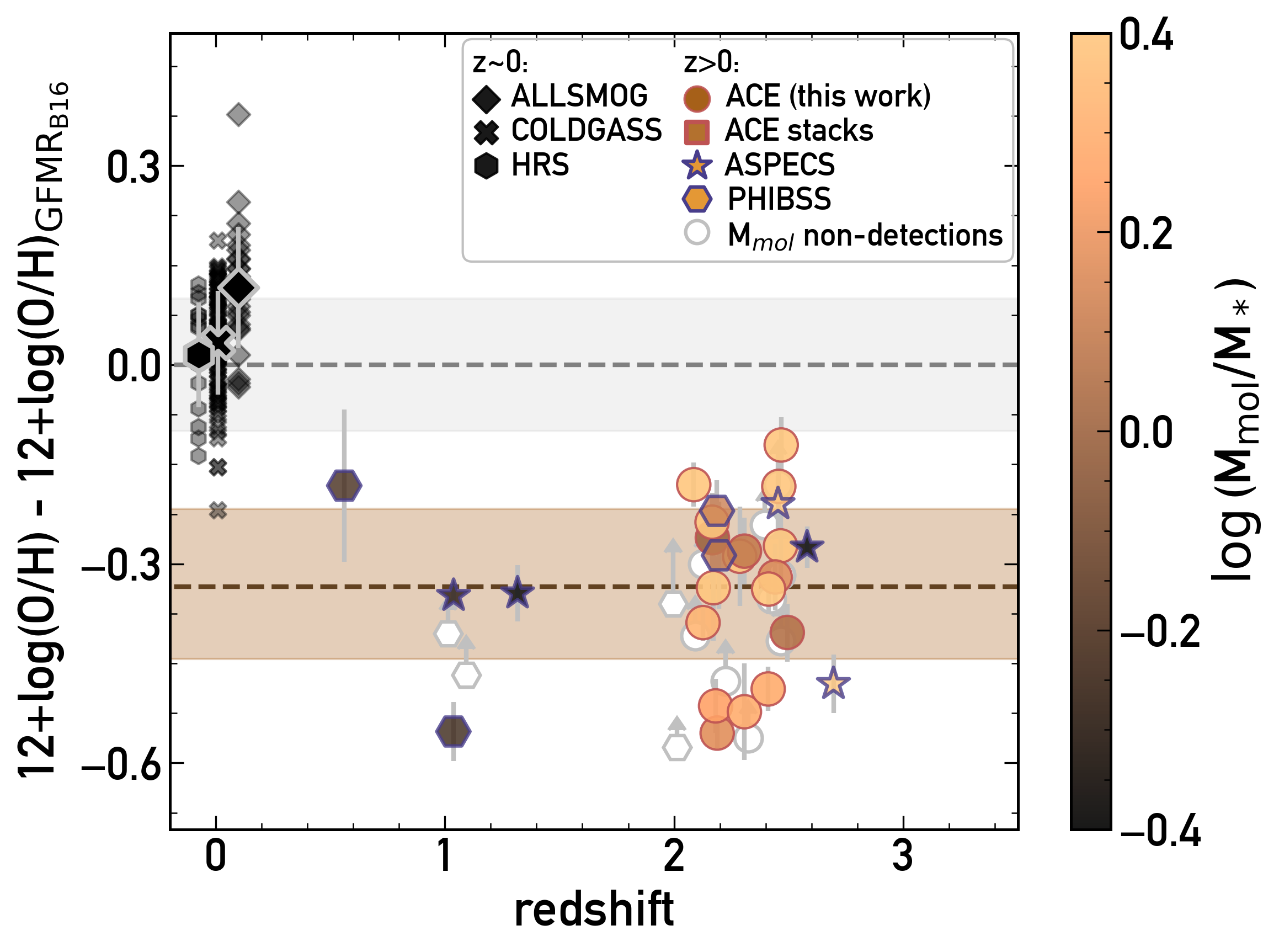}
    \caption{Comparing the measured metallicity and the predicted metallicity according to the local $\rm GFMR_{B16}$ (\citealt{Bothwell2016}), colour-coded by the molecular gas-to-stellar mass ratio. The average residuals of the local ALLSMOG (\citealt{Bothwell2014}), COLD GASS (\citealt{Saintonge2011}), and HRS (\citealt{Boselli2010}) surveys are shown with diamonds, crosses, and hexagons, respectively. The grey-shaded area shows a typical scatter of $0.1$ dex around GFMR$_{\rm B16}$. ASPECS and PHIBSS galaxies are shown with stars and hexagons, respectively. Non-detections in M$_{\rm mol}$ result in lower limits in the residuals, which we show with their corresponding markers. The dotted dark-brown line shows the average of our ACE residuals and the brown-shaded area shows the $1 \sigma$ standard deviation.}
    \label{fig:GFMR_redshift}
\end{figure}

In Fig.~\ref{fig:GFMR_redshift}, we show the residuals of this relation across $z\sim0-3$, defined as the difference between the observed metallicity and that predicted by $\rm GFMR_{ B16}$, colour-coded by $\mu_{\rm gas}$. We observe a clear $\sim0.3$ dex offset between the local Universe (diamond, cross and hexagon markers) and cosmic noon, as traced by the ACE sample (circles). To reconcile our measurements with $\rm GFMR_{B16}$, the inferred M$_{\rm mol}$ would need to be reduced by $\sim1$ dex. Adopting alternative $\alpha_{\rm CO}$, such as \citet{Bolatto2013b} instead of \citet{Accurso2017}, lowers M$_{\rm mol}$ by only $\sim0.5$ dex, which is insufficient to remove the observed offset in the GFMR residuals. This indicates that the uncertainties in $\alpha_{\rm CO}$ alone likely cannot account for the discrepancy. Another caveat is the adopted metallicity calibration. \citet{Bothwell2016} adopted an average of two metallicity tracers from \citet{Maiolino2008}, whereas we simultaneously used all available metallicity tracers from \citet{Sanders2025}, which is designed for the ISM conditions of high-redshift galaxies. To assess the impact of our adopted calibration, we recomputed our metallicities using \citet{Cataldi2025}, which is based on direct electron temperatures of cosmic noon SFGs. We find an average metallicity offset of $\sim0.04$ dex, substantially smaller than the $\sim0.3$ dex offset with respect to the local GFMR. This indicates that our conclusions are robust to the choice of calibration. We conclude that neither the uncertainties in $\alpha_{\rm CO}$ nor the choice of high-redshift metallicity calibration can account for the observed offset. Our results are therefore consistent with an evolution of the GFMR with cosmic time, suggesting stronger physical mechanisms lowering the metallicity of cosmic noon SFGs with respect to the ones acting in local SFGs (e.g. gas accretion and outflows). This result contrasts with the original interpretation of the FMR as redshift-invariant over $z\sim0-4$ (e.g. \citealt{Mannucci2010, MaiolinoMannucci2019, Curti2020, Langan2023}), although recent JWST results suggest that the FMR may break down at earlier epochs ($z>6$; e.g. \citealt{Curti2024}). In fact, we computed the FMR residuals using the local \citet{Curti2020} FMR and found that our sample remains around the zero line, i.e. our sample follows the redshift-invariant FMR. The observed deviation in the GFMR at $z\sim2$ may therefore indicate that physical processes, such as rapid metal dilution from gas accretion or efficient metal removal via outflows become important earlier when considering gas-regulated scaling relations.

Consistent with this picture, we find systematically higher gas fractions in the ACE sample ($\mu_{gas}\sim2$) compared to local galaxies ($\mu_{gas}\lesssim1$). While the cosmic evolution of the GFMR we observe might suffer from selection effects in the ACE sample -- CO detections become increasingly challenging at lower gas masses, so we are biased towards gas-rich reservoirs -- the elevated gas fractions observed in ACE are consistent with the typical properties of SFGs at cosmic noon and the global cosmic evolution of molecular gas reservoirs (e.g. \citealt{Tacconi2018}, \citealt{PerouxHowk2020}, \citealt{Walter2020}). The typical population of SFGs at cosmic noon is more gas-rich than their local counterparts; therefore, we consider the observed offset to reflect genuine evolution in the gas, star formation, and chemical properties of SFGs with cosmic time, although future observations probing lower $\mu_{gas}$ will be required to establish the degree to which sample selection contributes to the measured GFMR normalisation shift. We also note that the definition of $\mu_{gas}$ we adopt in this work assumes M$_{\rm mol}\gg$ other gas phases, i.e. if we could account for the total gas mass instead, $\mu_{gas}$ would be even larger. The offset we observe is therefore consistent with an earlier onset of the physical processes thought to drive the breakdown of the FMR, suggesting that different gas and star formation properties (i.e. M$_{\rm mol}$ vs SFR) may regulate chemical evolution on distinct timescales.

We note, however, that the ACE sample occupies a different region of parameter space than the sample of galaxies used to calibrate $\rm GFMR_{B16}$. While the stellar masses overlap, the ACE sample exhibits substantially higher SFRs than the local calibration samples. Therefore, part of our comparison relies on extrapolating the $\rm GFMR_{B16}$ relation beyond the regime in which it was originally calibrated, and the resulting residuals should be interpreted with some caution.

We also investigated the molecular gas depletion time, $t_{depl}=\rm M_{mol}/SFR$, which traces how long a galaxy would take to consume its molecular gas reservoir at the current SFR, and thus reflects the efficiency of star formation. Whereas the molecular gas–to–stellar mass, $\mu_{\rm gas}$, traces the overall gas richness of the system. At cosmic noon, galaxies typically exhibit both shorter depletion times and higher $\mu_{\rm gas}$ than their local Universe counterparts (e.g. \citealt{Tacconi2018}), implying more efficient star formation and larger gas reservoirs. This is also seen in our ACE sample (\citealt{Popping+}). In Fig.~\ref{fig:GFMR_GFMRres} (right panel), we tentatively find that, at a fixed M$_*$, galaxies with higher $\mu_{\rm gas}$, i.e. more gas-rich, show stronger negative metallicity offsets. This trend is consistent with a scenario in which high-redshift galaxies are continuously supplied with metal-poor gas: such inflows both increase the gas fraction and dilute the ISM, while simultaneously sustaining elevated SFRs and the corresponding short depletion times. 

\subsection{A word of caution}
\label{subsec:caveats}
A key choice in any work using CO data to derive M$_{\rm mol}$ is the adoption of a suitable $\alpha_{\text{CO}}$ conversion function. Using the dynamical mass method (e.g. S23), we investigated a metallicity-dependent $\alpha_{\text{CO}}$ at subsolar metallicities in a representative sample of SFGs, leveraging homogeneous metallicity and CO line measurements, for the first time. We note that the dynamical mass method carries significant uncertainties, including uncertainties on M$_*$ measurements, assuming negligible atomic gas M$_{\rm HI}$ (e.g. \citealt{Bigiel2016}, \citealt{Saintonge2017}) and dark matter (e.g. \citealt{Genzel2020}, \citealt{Price2021}) within galaxy discs. Despite uncertainties, our results support a metallicity dependence consistent with literature calibrations. These calibrations, however, can span over 1 dex in $\alpha_{\text{CO}}$ at fixed metallicities, reflecting differences in metallicity scales (e.g. \citealt{PettiniPagel2004}, \citealt{PilyuginThuan2005}), diagnostics (e.g. O3N2, R23), and sample selection which introduces significant systematics. In the absence of metallicity measurements, the community often adopts fixed values such as $\alpha_{\text{CO}}\sim0.8$ for starbursts or a Milky-Way $\alpha_{\text{CO}}\sim3.6$ (\citealt{Bolatto2013b}) for main-sequence SFGs (e.g. \citealt{Daddi2010}, \citealt{Aravena2019}), highlighting the dependence of $\alpha_{\text{CO}}$ on galaxy physical conditions. This choice significantly impacts the derived M$_{\rm mol}$ and subsequent results depending on such M$_{\rm mol}$.

In this work, we took advantage of the current observational constraints -- leveraging a uniquely complete dataset that combines CO and metallicity measurements for the same sample of typical SFGs -- to constrain $\alpha_{\text{CO}}$ at subsolar metallicities. While this represents a step forward, significant uncertainties remain. As seen in the left panel of Fig.~\ref{fig:alphaCO}, at $12+\log \rm O/H < 8.69$, where our sample lies, the different literature calibrations show a spread up to $\sim 2$ dex, resulting in M$_{\rm mol}$ differing by up to two orders of magnitude as well, depending on the choice of calibration. One of the largest uncertainties in our $\alpha_{\rm CO}$ measurements comes from the derivation of the dynamical mass. Improving the accuracy of $\alpha_{\text{CO}}$  will ultimately require resolved CO kinematics to accurately measure rotation curves. This effort would allow more appropriate $\alpha_{\text{CO}}$ calibrations for different galaxy populations across cosmic epochs. Nonetheless, uncertainties in measuring M$_{\rm mol}$ will remain even with the highest CO data quality, simply because of physical limitations imposed by CO-dark gas (e.g. \citealt{Bisbas2015}, \citealt{Madden2020}). This limitation can only be alleviated by using multiple independent tracers, such as atomic carbon [CI], or dust-continuum, to robustly measure M$_{\rm mol}$.

\section{Summary and conclusions}
\label{section:five}
In this work, we presented a self-consistent analysis of the cold molecular gas, stellar and metal content of typical subsolar SFGs at cosmic noon, using data from the ALMA Large Programme ACE. Our findings are summarised as follows:

\begin{itemize}
    \item We investigated the GFMR, a three-parameter relation between gas-phase metallicity, M$_*$, and M$_{\rm mol}$. Within our sample, we find no statistically significant evidence of secondary dependence of metallicity on M$_{\rm mol}$ as traced by CO emission. The intrinsic scatter of the GFMR $\sigma_{\rm GFMR}\sim0.11$ is comparable to that of the FMR, ($\sigma_{\rm FMR}\sim0.13$).
    \item We observe a tentative trend of chemical enrichment following lower M$_{\rm mol}/\rm M_*$, suggesting replenishment of the molecular gas reservoir and enrichment of the ISM via star formation occurring on different timescales.
    \item Our observational findings at cosmic noon are in agreement with theoretical gas-regulator models describing the evolution of the metallicity within the ISM as driven by gas flows, in particular gas outflows. We find that an average of outflow mass loading factors $\eta_{\rm out} \sim4$, i.e. subsolar SFGs at cosmic noon are efficient at removing mass and metals via molecular outflows.
    \item We compared the metallicity of our sample to that predicted by the local \citet{Bothwell2016} $\rm GFMR_{B16}$) and find that $\rm GFMR_{B16}$) over-predicts the metal content by 0.3 dex. The observed discrepancy between local samples and ours at cosmic noon likely comes from the higher M$_{\rm mol}/\rm M_*$, reaffirming the interpretation of the different timescales between the evolution of the molecular gas reservoir and the chemical enrichment.
    \item Our data favours the $\alpha_{\rm CO}$ prescription of \citet{Accurso2017}, while remaining broadly consistent with other literature calibrations at subsolar metallicities. Dedicated spectroscopic observations are required to more precisely constrain $\alpha_{\rm CO}$ across metallicities.
\end{itemize}

This work provides the first exploration of the interplay between chemical enrichment and molecular gas content at cosmic noon, in a well-controlled sample of typical SFGs. This sample additionally allows us to probe an uncharted parameter space, reaching subsolar metallicities. Our findings demonstrate that subsolar SFGs at cosmic noon are already in a well-regulated state in terms of their baryon cycle and chemical enrichment, albeit the statistics and uncertainties of our observations. In future work we will search for more direct evidence of molecular outflows, such as via high-resolution CO observations or stacking techniques, to further test gas-regulator models predicting strong molecular outflows capable of ejecting mass and metals. Other paths to explore thanks to this unique ACE dataset include using other molecular gas tracers (e.g. atomic carbon [CI] or the existing ACE dust continuum data) to probe potential CO-dark molecular gas, or use high spatially resolved multi-wavelength data to study the distribution of the chemical enrichment with respect to the distribution of the molecular gas.

\begin{acknowledgements}
This paper makes use of the following ALMA data: ADS/JAO.ALMA\#2018.1.01128.S, 2024.1.00534.L. ALMA is a partnership of ESO (representing its member states), NSF (USA) and NINS (Japan), together with NRC (Canada), MOST and ASIAA (Taiwan), and KASI (Republic of Korea), in cooperation with the Republic of Chile.
We acknowledge assistance and computational support provided by Allegro, the European ALMA Regional Center node in the Netherlands. This work has been funded by the Atracción de Talento Grant No. 2022-T1/TIC-20472 of the Comunidad de Madrid, Spain, and the European Research Council (ERC) under the European Union’s Horizon 2020 research and innovation programme (DistantDust, Grant agreement No. 101117541). L.A.B. acknowledges support from the Dutch Research Council (NWO) under grant VI.Veni.242.055 (\url{https://doi.org/10.61686/LAJVP77714}). LA acknowledges support by the CSIC Program 'Programa JAE' (JAE-Pre 2023), by the grant PID2024-158856NA-I00 funded by Spanish Ministerio de Ciencia e Innovación MCIN/AEI/10.13039/501100011033 and by “ERDF A way of making Europe”. DN is grateful for support from NASA via grants ATP-21-0013 and ATP-23-0002. MP is funded by NASA grant ATP-23-0002.
\end{acknowledgements}

\bibliographystyle{aa} 
\bibliography{biblio.bib}

\FloatBarrier
\begin{appendix}
\section{Additional material}
\label{section:apd}

\begin{figure}[!ht]
    \centering
    \includegraphics[width=0.9\columnwidth]{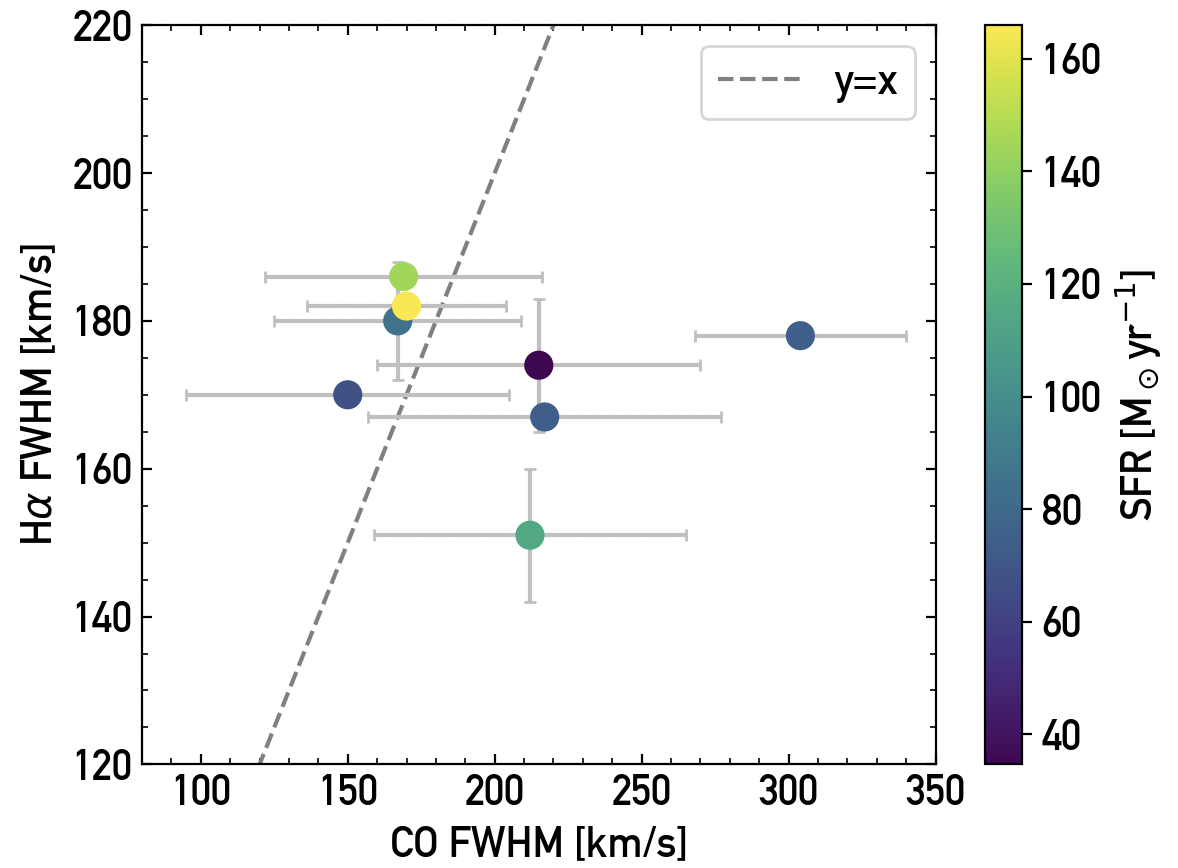}
    \caption{FWHM of the H$\alpha$ emission line measured in the MOSDEF survey (\citealt{Kriek2015}; \citealt{Reddy2015}) and the FWHM of the CO(3--2) emission line measured in ACE. We only show the galaxies for which we ran a free fit (see Sect.~\ref{subsubsec:spectra}). The 1:1 relation is shown with the dashed grey line.}
    \label{fig:COHAFWHM}
\end{figure}

\begin{table}[!ht]
\caption{Physical properties of our sample, as presented in \citet{Shivaei+}.}
\label{table:pties}
\centering
\small
\begin{tabular}{|ccccc|}
\hline
\toprule
\textbf{ID} &  \textbf{Redshift} & \textbf{12+log$ ({\rm O/H})$} & \textbf{logM$_*$} & \textbf{logSFR}\\
(1) & (2) & (3) & (4) & (5)\\
\hline
\midrule
2672& 2.3074 & $8.42_{-0.06}^{+0.05}$ & $10.61_{-0.19}^{+0.11}$ & $1.87 \pm 0.01$\\

3324& 2.3072 & $8.61_{-0.04}^{+0.04}$& $10.30_{-0.08}^{+0.08}$ & $1.93 \pm 0.19$\\

3626& 2.3247 & $8.27_{-0.03}^{+0.03}$& $10.17_{-0.05}^{+0.04}$ & $1.33 \pm 0.04$\\

3666 & 2.0859 & $8.36_{-0.02}^{+0.02}$& $9.20_{-0.06}^{+0.10}$ & $1.92 \pm 0.05$\\

3773& 2.4251 & $8.32_{-0.03}^{+0.03}$& $9.64_{-0.07}^{+0.07}$ & $1.75 \pm 0.12$\\

4497&2.4414 & $8.56_{-0.04}^{+0.04}$& $10.25_{-0.11}^{+0.09}$ & $1.83 \pm 0.13$\\

5094 &2.1715 & $8.53_{-0.05}^{+0.05}$& $10.33_{-0.27}^{+0.13}$ & $1.85 \pm 0.15$\\

5814&2.1266 & $8.49_{-0.02}^{+0.02}$& $10.34_{-0.05}^{+0.05}$ & $2.06 \pm 0.07$\\

5901& 2.3962 & $8.48_{-0.02}^{+0.02}$& $9.72_{-0.06}^{+0.05}$ & $1.63 \pm 0.06$\\

6283&2.2238 & $8.26_{-0.03}^{+0.03}$&$9.85_{-0.11}^{+0.09}$ & $1.69 \pm 0.04$\\

6750&2.1271 & $8.40_{-0.02}^{+0.02}$&$9.67_{-0.06}^{+0.05}$ & $1.74 \pm 0.05$\\

8280&2.4941 & $8.56_{-0.05}^{+0.04}$&$10.59_{-0.02}^{+0.02}$ & $1.75 \pm 0.20$\\

8515&2.4537 & $8.32_{-0.03}^{+0.03}$&$9.11_{-0.07}^{+0.16}$ & $1.83 \pm 0.07$\\

9393&2.4127 & $8.43_{-0.02}^{+0.02}$&$9.89_{-0.09}^{+0.10}$ & $1.90 \pm 0.06$\\

9971&2.4108 & $8.41_{-0.02}^{+0.02}$&$10.41_{-0.08}^{+0.08}$ & $1.86 \pm 0.04$\\

13296&2.1672 & $8.63_{-0.05}^{+0.05}$&$10.26_{-0.06}^{+0.23}$ & $1.54 \pm 0.16$\\

13701&2.1659 & $8.59_{-0.03}^{+0.02}$&$10.12_{-0.02}^{+0.06}$ & $2.22 \pm 0.09$\\

16594 &2.2863 & $8.50_{-0.03}^{+0.03}$&$9.96_{-0.07}^{+0.07}$ & $1.43 \pm 0.10$\\

19013& 2.4571 & $8.58_{-0.03}^{+0.02}$&$9.87_{-0.07}^{+0.07}$ & $1.87 \pm 0.24$\\

19439&2.4663 & $8.29_{-0.02}^{+0.03}$& $9.75_{-0.09}^{+0.08}$ & $1.92 \pm 0.07$\\

%19753 & 20661 & 2.4694& $158\pm 34$ & $1.87 \pm 0.01$\\

19985&2.1882 & $8.37_{-0.01}^{+0.01}$&$10.45_{-0.06}^{+0.05}$ & $2.16 \pm 0.03$\\

21955&2.4676 & $8.55_{-0.03}^{+0.03}$&$9.62_{-0.10}^{+0.09}$ & $1.86 \pm 0.10$\\

22193&2.4651 & $8.46_{-0.03}^{+0.03}$& $9.89_{-0.08}^{+0.15}$ & $1.76 \pm 0.19$\\
 
24020&2.0923 & $8.38_{-0.03}^{+0.03}$& $9.95_{-0.09}^{+0.07}$ & $1.46 \pm 0.13$\\

24763&2.4644 & $8.50_{-0.05}^{+0.05}$&$9.96_{-0.07}^{+0.06}$& $2.16 \pm 0.23$\\

25229&2.1813 & $8.34_{-0.03}^{+0.03}$&$10.20_{-0.08}^{+0.07}$ & $1.38 \pm 0.09$\\[3pt]

\hline
\bottomrule
\end{tabular}
\tablefoot{(1) lists the ID from the 3D-HST v4 catalogue. (2) is the redshift as measured from multiple optical emission lines from MOSFIRE spectra (\citealt{Kriek2015}). (3) is the metallicity measured from multiple optical/NIR emission lines using strong line calibrations from \citet{Sanders2025}. (4) is the M$_*$ in M$_\odot$, derived via \texttt{Prospector} SED fitting. (5) is the SFR in M$_\odot$ yr$^{-1}$, measured from dust-corrected H$\alpha$ emission line.}
\end{table}

\begin{table}[htbp]
\caption{Best-fit parameters of the CO(3--2) emission line for which the line was detected in our sample. }
\label{table:Fitparams}
\centering
\begin{tabular}{|ccccc|}
\hline
\toprule
\textbf{ID} & \textbf{Peak} & \textbf{Centre} & \textbf{FWHM} & \textbf{Fit}\\
(1) & (2) & (3) & (4) & (5)\\
\hline
\midrule
2672 & $0.47\pm 0.11$ & $77\pm 50$ & $304\pm 36$ & 1\\

3324 & $0.50\pm 0.10$ & $-22\pm 18$ & $167\pm 42$ & 1\\

3666 & $0.28\pm 0.06$ & $-79\pm 19$ & ($138\pm21$) & 0\\

4497 & $0.39\pm 0.15$ & $-59\pm 31$ & $150\pm55$ & 1\\

5094 & $0.69\pm 0.13$ & $-97\pm 38$ & ($284\pm43$) & 0\\

5814 & $0.52\pm 0.09$ & $2\pm 16$ & $212\pm 53$ & 1\\

8280 & $0.72\pm 0.08$ & $-26\pm9$ & $151\pm17$ & 1\\

9393 & $0.30\pm 0.07$ & $-47\pm 11$ & ($120\pm23$) & 0\\

9971 & $0.37\pm 0.09$ & $33\pm 34$ & ($216\pm23$) & 0\\

13296 & $0.40\pm 0.08$ & $8\pm23$ & $215\pm55$ & 1\\

13701 & $0.67\pm 0.12$ & $37\pm22$ & $170\pm34$ & 1\\

16594 & $0.34\pm 0.07$ & $-5\pm21$ & ($183\pm22$) & 0\\

19013 & $0.30\pm 0.06$ & $-47\pm34$ & ($221\pm23$) & 0\\

%19753 & 20661 & $0.73\pm 0.17$ & $62\pm33$ & $217\pm60$ & 1\\

19985 & $0.34\pm 0.08$ & $-40\pm29$ & $169\pm47$ & 1\\

21955 & $0.33\pm 0.07$ & $11\pm31$ & ($207\pm23$) & 0\\

24763 & $0.29\pm 0.05$ & $-27\pm29$ & ($223\pm23$) & 0\\

25229 & $0.25\pm 0.07$ & $80\pm28$ & ($138\pm22$) & 0\\[3pt]
\hline
\bottomrule
\end{tabular}
\tablefoot{(1) lists the IDs from the 3D-HST v4 catalogue. (2) lists the peak of the Gaussian fit in millijansky. (3) lists the centre of the Gaussian fit in kilometres per second. (4) lists the FWHM of the Gaussian fit in in kilometres per second. (5) indicates whether the Gaussian fit of the CO(3--2) spectrum is fully free (`1') or constrained to the H$_\alpha$ FWHM (`0').}
\end{table}

\begin{figure}[htbp]
    \centering
    \includegraphics[width=\columnwidth]{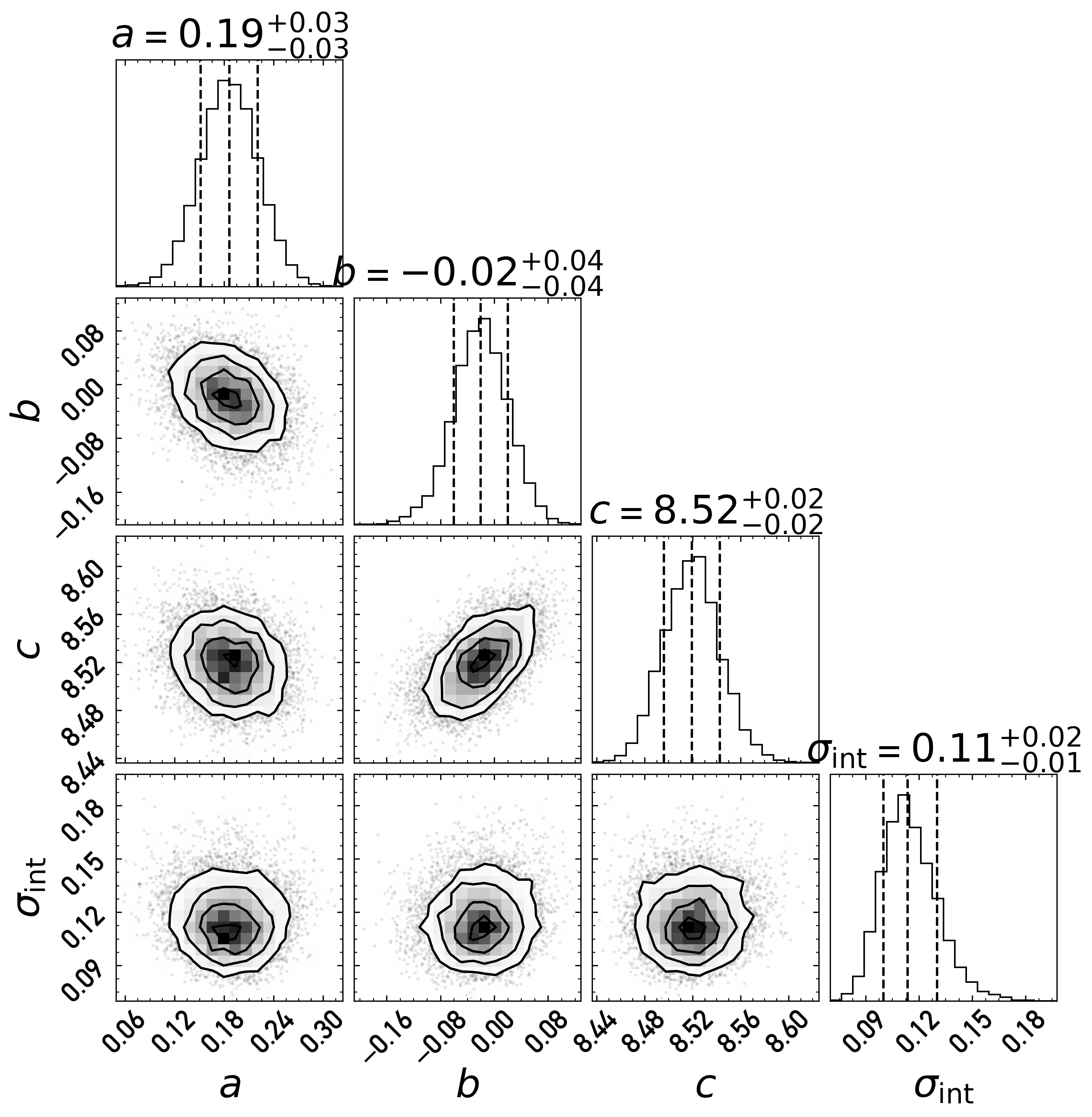}
    \caption{Posterior distributions of the GFMR linear fit using \texttt{PyMC}. $a$ represents the dependence on M$_*$; $b$ represents the dependence on M$_{\rm mol}$; $c$ represents the intercept; and $\sigma_{\rm int}$ represents the intrinsic scatter about the relation, accounting for observational errors.}
    \label{fig:cornerplot}
\end{figure}

\begin{figure*}[]
    \centering
    \includegraphics[width=\textwidth]{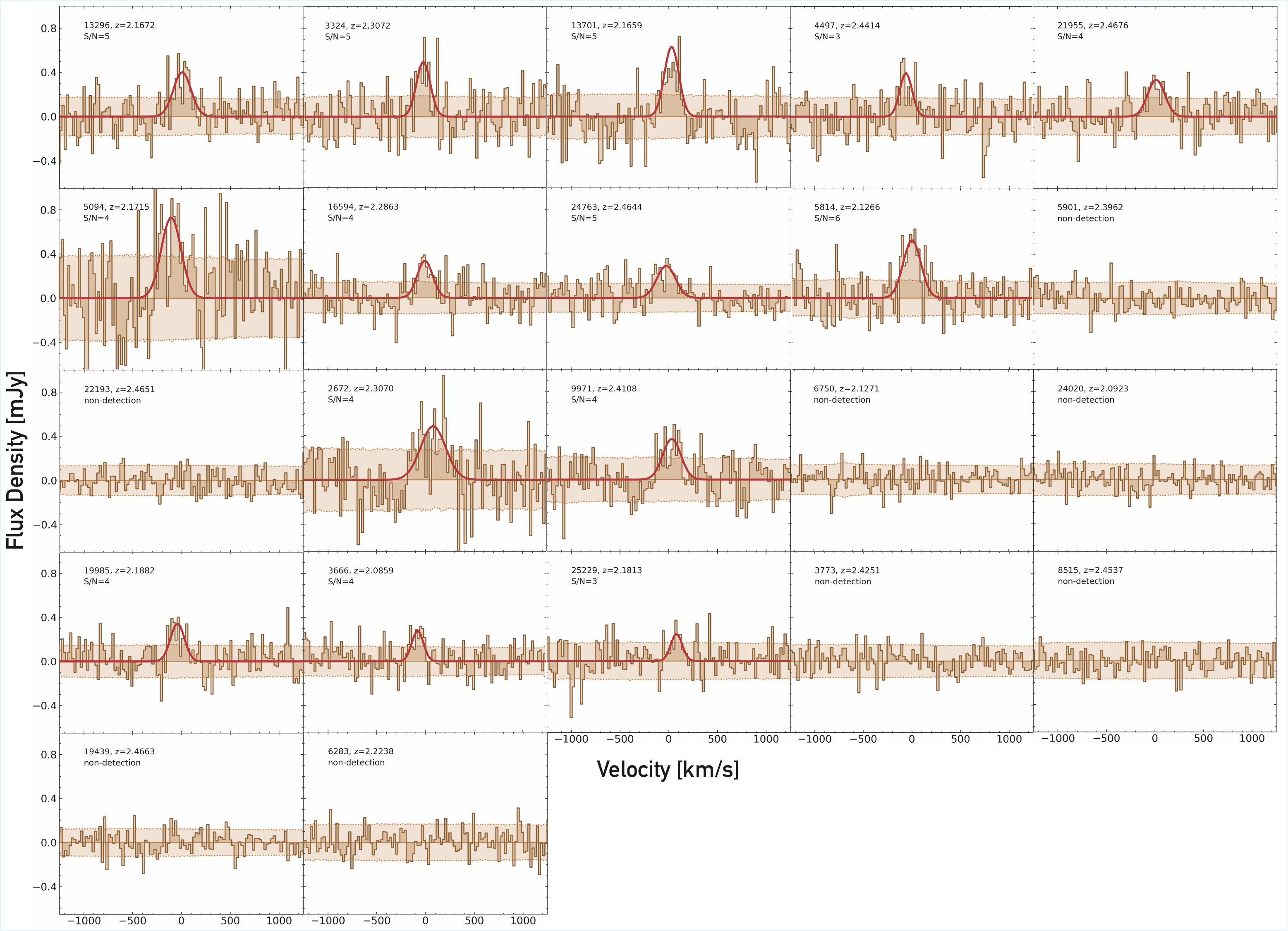}
    \caption{ACE CO(3--2) spectra and their best Gaussian fit in the case of detections (thick red line), ordered by decreasing metallicity. All subplots show the target IDs and their redshifts in the upper left corner. We also show the rms per channel with the horizontal beige-shaded area within the dotted lines.}
    \label{fig:spectra}
\end{figure*}

\begin{table*}[]
\caption{Physical properties of the literature sample included in our analysis from the ASPECS (upper block; \citealt{Walter2016}) and PHIBSS (lower block; \citealt{Tacconi2018}) surveys.}
\label{table:ptiesliterature}
\centering
\begin{tabular}{|cccccc|}
\hline
\toprule
\textbf{ID} &  \textbf{Redshift} & \textbf{12+log$({\rm O/H})$} & \textbf{logM$_*$} & \textbf{logSFR} & \textbf{logM$(_{\rm mol})$}\\
(1) & (2) & (3) & (4) & (5) & (6)\\
\hline
\midrule         
ASPECS/1mm.04& 2.454 & $8.65_{-0.03}^{+0.033}$& $10.33_{-0.09}^{+0.09}$ & $1.76 \pm 0.05$ & $10.85_{-0.06}^{+0.05}$\\
ASPECS/1mm.06& 2.696 & $8.54_{-0.04}^{+0.04}$& $10.99_{-0.06}^{+0.05}$ & $2.12 \pm 0.05$ & $11.42_{-0.08}^{+0.08}$\\
ASPECS/1mm.07& 2.581 & $8.16_{-0.03}^{+0.03}$& $10.91_{-0.03}^{+0.03}$ & $1.43 \pm 0.09$ & $10.3_{-0.1}^{+0.1}$\\
ASPECS/1mm.13& 1.038 & $8.65_{-0.01}^{+0.01}$& $10.64_{-0.03}^{+0.04}$ & $1.36 \pm 0.06$ & $10.34_{-0.06}^{+0.06}$\\
ASPECS/1mm.14$^\dag$& 1.997 & $8.71_{-0.04}^{+0.04}$& $10.7_{-0.3}^{+0.2}$ & $0.4 \pm 0.9$ & $<9.6$\\
ASPECS/1mm.16& 1.317 & $8.84_{-0.03}^{+0.02}$& $11.13_{-0.08}^{+0.08}$ & $-0.3 \pm 0.9$ & $10.0_{-0.3}^{+0.3}$\\
ASPECS/1mm.21$^\dag$& 1.093 & $8.64_{-0.04}^{+0.04}$& $10.82_{-0.03}^{+0.04}$ & $0.86 \pm 0.07$ & $<9.8$\\[3pt]
\hline
EGS13017707& 1.037 & $8.57_{-0.02}^{+0.02}$&$11.21_{-0.13}^{+0.10}$ & $2.10\pm0.04$& $10.95_{-0.06}^{+0.05}$\\
GN4-29743&2.1873&$8.67_{-0.02}^{+0.02}$&$10.30_{-0.09}^{+0.13}$ & $1.9\pm0.1$& $10.4_{-0.2}^{+0.2}$\\
GN4-1964-L14GN033& 0.561 & $8.70_{-0.13}^{+0.09}$&$10.14_{-0.04}^{+0.04}$ & $0.7\pm0.1$& $9.9_{-0.4}^{+0.4}$\\
GN4-19913$^\dag$&2.0128&$8.67_{-0.02}^{+0.02}$&$11.81_{-0.02}^{+0.01}$ & $2.7\pm0.1$& $<11.6$\\ 
GN4-7054$^\dag$& 1.013 & $8.82_{-0.01}^{+0.02}$&$11.32_{-0.07}^{+0.13}$ & $1.7\pm0.1$ & $<10.2$\\
zC406690& 2.196 & $8.69_{-0.04}^{+0.04}$&$10.7_{-0.2}^{+0.2}$ & $2.5\pm0.5$& $10.8_{-0.2}^{+0.2}$\\[3pt]
\hline
\bottomrule
\end{tabular}
\tablefoot{(1) lists the object ID. (2) lists the redshift as measured from interferometric spectroscopy data. (3) lists the metallicity measured from multiple strong optical/NIR emission lines using \citet{Sanders2025}. (4) and (5) list the stellar mass (M$_*$) in solar masses and the SFR in solar masses per year, respectively. Both are derived from \texttt{Prospector} SED fitting. (6) lists the M$_{\rm mol}$ in solar masses, derived using the lowest detected CO J transition and the \citet{Accurso2017} metallicity-dependent $\alpha_{\rm CO}$. We report 3$\sigma$ upper limits for sources marked with $^\dag$.}
\end{table*}
\end{appendix}
\end{document}